\documentclass[11pt,a4paper]{article}
\usepackage{jcappub}
\usepackage{bm}
\usepackage{amsmath}
\usepackage{amssymb}
\usepackage{graphicx}
\usepackage{hyperref}
\usepackage{color}
\usepackage{mathtools}
\usepackage{times}
\usepackage{hyperref}
\usepackage{comment}
\hypersetup{
    colorlinks=true,
    linkcolor=blue,
    filecolor=magenta,
    citecolor=blue,
    urlcolor=blue,
}

\newcommand{\up}[1]{{\rm #1}}
\newcommand{\bdv}[1]{{\bf #1}}
\newcommand{\beeq}{\begin{equation}}
\newcommand{\eneq}{\end{equation}}
\newcommand{\pa}{\partial}
\newcommand{\Dquad}{\qquad\qquad}
\newcommand{\fnl}{f_\up{NL}}
\newcommand{\OO}{\mathcal{O}}
\newcommand{\Vang}{\bdv{\hat n}}
\newcommand{\de}{\delta}
\newcommand{\HH}{\mathcal{H}}   
\newcommand{\LL}{\mathcal{L}}   

\begin{document}

\begin{titlepage}

\setcounter{page}{1} \baselineskip=15.5pt \thispagestyle{empty}
\pagenumbering{roman}

\bigskip

\vspace{1cm}
\begin{center}
{\fontsize{20}{28}\selectfont \bfseries
Incompatibility of Standard Galaxy Bias Models in General Relativity}
\end{center}

\vspace{0.2cm}

\begin{center}
{\fontsize{13}{30}\selectfont Jaiyul Yoo$^{a,b}$}
\end{center}

\begin{center}
\vskip 8pt
\textsl{$^a$ Center for Theoretical Astrophysics and Cosmology,
Institute for Computational Science}\\
\textsl{University of Z\"urich, Winterthurerstrasse 190,
CH-8057, Z\"urich, Switzerland}

\vskip 7pt

\textsl{$^b$Physics Institute, University of Z\"urich,
Winterthurerstrasse 190, CH-8057, Z\"urich, Switzerland}

\vskip 7pt

\today

\end{center}

\note{jyoo@physik.uzh.ch}

\vspace{1.2cm}
\hrule \vspace{0.3cm}
\noindent {\sffamily \bfseries Abstract} \\[0.1cm]
The standard model for galaxy bias is built in a Newtonian framework,
and several attempts have been made in the past to put it in a relativistic
framework. The focus of past works was, however, to use
the same Newtonian formulation, but to provide its interpretation 
in a relativistic framework
by either fixing a gauge condition or transforming to a local coordinate
system. Here we demonstrate that these reverse-engineered approaches 
 do not respect the diffeomorphism symmetry in general relativity, and
we need to develop a covariant model of galaxy bias that
is diffeomorphism compatible. We consider a simple toy model
for galaxy bias and discuss the impact for measuring
the primordial non-Gaussianity. 
\vskip 10pt
\hrule

\vspace{0.6cm}
\end{titlepage}

\clearpage

\noindent\hrulefill \tableofcontents \noindent\hrulefill

\pagenumbering{arabic}

\section{Introduction}
Luminous galaxies provide a powerful way to map the
large-scale structure of the Universe, and the relation between the galaxy
and the matter distributions is known as the galaxy bias 
(\cite{PRSC74,KAISE84,POWI84,BBKS86,BCEK91}, see also
\cite{DEJESC18} for a recent review). In the simplest model 
\cite{PRSC74,KAISE84,POWI84,BBKS86,BCEK91}, 
the fluctuation $\de_g:=\de n_g/\bar n_g$
in the galaxy number density~$n_g$ is related to the matter
density fluctuation~$\de_m:=\de\rho_m/\bar\rho_m$ as
\beeq
\label{linear}
\de_g=b~\de_m~,
\eneq
where the galaxy bias factor~$b$ is a constant. The linear bias relation
in eq.~\eqref{linear} has been widely used to extract cosmological information
on large scales (e.g., \cite{EIZEET05,TEEIET06,ANAUET12,AUBAET15}),
and extensive efforts in the community have been
devoted to building a more precise model for galaxy bias on nonlinear scales
\cite{BOMY96,MOWH96,MATSU99,CAPOKA00,MAFR00,SELJA00,SMT01,SCSHET01,BEWE02,
COSH02,MCDON06,MCRO09,BASEET12,DESJA13,SCJEDE13,SENAT15,MISCZA15,ANFAET15,
IVSIZA20,DAGLET20}. So far, most of the approaches to modeling galaxy bias are built
in a Newtonian framework. Since measurement precision in large-scale surveys
is highest on small scales, where the Newtonian dynamics is accurate,
it is natural and practical that most resources in terms of observational
and theoretical investigations are devoted in this direction.

However, viewed from a relativistic perspective, 
the simple linear bias relation in eq.~\eqref{linear} is ambiguous,
because the matter density fluctuation gauge-transforms as
\beeq
\tilde \de_m=\de_m+3\HH~T+\OO(2)~,
\eneq
under a general coordinate transformation $\tilde x^\mu=x^\mu+\zeta^\mu$,
where the infinitesimal transformation is $\zeta^\mu=:(T,\LL^i)$,
the conformal Hubble parameter is~$\HH=a'/a$, and the prime denotes
the derivative with respect to the conformal time~$\eta$. Similarly,
the galaxy number density fluctuation also gauge-transforms as
\beeq
\label{galaxy}
\tilde\de_g=\de_g-{\bar n_g'\over\bar n_g}~T+\OO(2)~.
\eneq
Therefore, if we want to maintain the linear bias relation in eq.~\eqref{linear}
in all coordinates, the evolution of the background 
galaxy number density~$\bar n_g$ should be related
to the bias parameter as
\beeq
\label{evol}
{d\ln \bar n_g\over d\eta}=-3\HH ~b~,\Dquad\up{or}\Dquad
{d\ln \bar n_g\over d\ln(1+z)}=3~b~,
\eneq
which we already know is {\it not true} in reality
(see, e.g., \cite{EIANET01,WHBLET11} for recent observations).
In contrast, note that eq.~\eqref{evol} is perfectly valid for the case of 
the matter density as a model for galaxy, i.e.,
$n_g\equiv\rho_m$ with $b=1$. Furthermore, the relation
$\rho_m=\bar\rho_m(1+\de_m)$ is valid in all coordinates or covariant.
This illustrates the issues that the standard model for galaxy bias
is {\it incompatible with the diffeomorphism symmetry} in general relativity.

Since galaxy clustering is essentially 
the dynamics of non-relativistic
matter components, the contributions of the relativistic effects in galaxy
clustering are naturally
 small compared to the matter density fluctuations, unless
the scales of interest are close to the horizon scales \cite{YOO10,YOHAET12}.
In this respect,
the nature of the current investigation is more theoretical than of
practical applications today,
 and the standard approaches to modeling galaxy bias 
work just fine in most cases.
However, the primordial non-Gaussian signal~$\fnl$ that
might have been imprinted in the early Universe exhibits the strongest
signals in galaxy clustering
on scales close to the horizon scales \cite{DADOET08,SLHIET08,MAVE08}, 
and its measurements
in galaxy clustering are one of the main goals in the
upcoming large-scale surveys such as 
\cite{LSST04,EUCLID11,WFIRST12,ANARET18,MUREET22,CAIVET22,DALEET22}
with the target
uncertainty $\Delta\fnl\approx$ few.
In particular, the relativistic
effects on such large scales are shown to affect the inference of the
primordial non-Gaussian signal from the measurements
\cite{BRHIET14,BRHIWA14,VIVEMA14,CAMASA15,MAPIRO21,CADI22,FOPAET23},
and hence a correct
modeling of galaxy bias in a relativistic framework is the last missing piece
toward a fully relativistic description of galaxy clustering 
\cite{MIYO20,YOGRMI22}.

Here we critically investigate
 the issues in the standard approaches to modeling galaxy
bias from a relativistic perspective and discuss ways to build galaxy
bias models that are compatible with the diffeomorphism symmetry in
general relativity.

\section{Proper-time hypersurface for modeling galaxy bias} 
\subsection{Gauge choice for galaxy bias: temporal and spatial} 
\begin{figure}[t]
\centering
\includegraphics[width=0.95\textwidth]{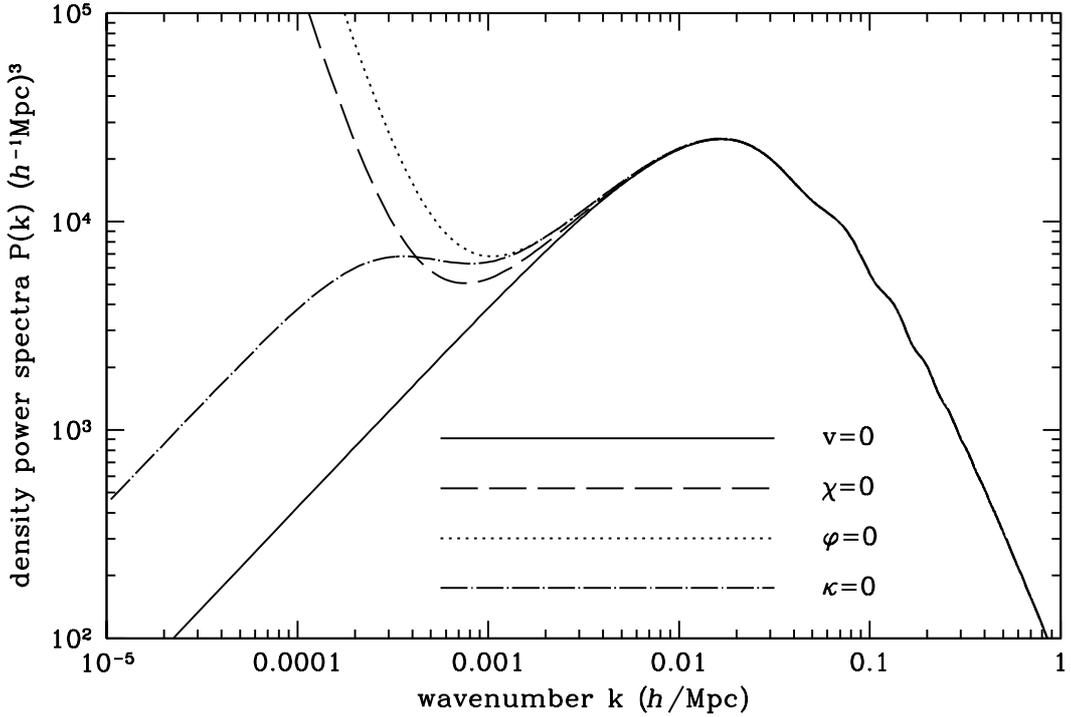}
\caption{Matter density fluctuations in various gauges. There exist an
infinite number of possibilities for a gauge choice, though the conformal
Newtonian gauge (dashed: $\chi\equiv0$) and the comoving-synchronous gauge 
(solid: $v\equiv0$) are most popular in literature.
Shown in the plot are two additional  gauge 
choices: uniform curvature (dotted: $\varphi\equiv0$) and uniform expansion
(dot dashed: $\kappa\equiv0$). 
While the differences in the matter density fluctuations lie mostly
on large scales, one can choose a gauge condition 
to make it vanish completely, i.e., uniform density gauge ($\de_m\equiv0$). 
The notation for metric in the plot follows the convention
described in Appendix \cite{YOGRET18}, and for the plot
we assumed the best-fit cosmological
parameter set for a $\Lambda$CDM model from the Planck measurements
\cite{PLANCKcos18}.}
\label{matter}
\end{figure}

Since observers
can provide unique numbers for~$\de_g^\up{obs}$ in any given surveys,
theorists need to predict unique numbers for~$\de_g$ in any given models
to compare to~$\de_g^\up{obs}$. Given the freedom to choose any coordinate
system in general relativity, a simple way to remove the ambiguities
shown in eq.~\eqref{galaxy} due to gauge choice and at the same time
to maintain the linear bias
relation in eq.~\eqref{linear} as in the Newtonian dynamics is to fix
a gauge for the matter density fluctuation~$\de_m$ in eq.~\eqref{linear} 
{\it by hand}. The immediate question then arises ``{\it which gauge choice for 
the theoretical prediction of~$\de_m$ in modeling galaxy
bias?}''  Figure~\ref{matter} illustrates 
that a different choice of gauge condition leads to a different prediction
for~$\de_m$ (hence~$\de_g$), which demands 
that a choice of gauge condition needs 
to be made and physically justified for the linear bias relation in 
eq.~\eqref{linear} to be valid.

The synchronous gauge was chosen \cite{CHLE11} for the linear bias relation
in the relativistic dynamics,
and this choice of gauge condition was supported \cite{JESCHI12,JESC13} by
physical arguments that the only clock available for local observers
in galaxies is the proper time along the world line of the local observers
in the galaxies.
The proper-time hypersurface of those
observers moving along a time-like geodesic 
coincides with
those in the synchronous gauge and in the matter-comoving gauge at the linear
order in perturbations \cite{YOO14b}, though two gauge conditions are 
different. The physical arguments for the proper-time hypersurface in
 modeling galaxy bias are sound and indeed compatible with the diffeomorphism
symmetry in general relativity, as the four-velocity~$u^\mu$ of the galaxies
defines the time-direction for the local observers in the rest frame of
the galaxies and a hypersurface orthogonal to the four-velocity defines
the proper-time hypersurface in a covariant manner. In this way, the
temporal gauge choice for modeling galaxy bias
can be made, {\it not} by a specific hand-picked gauge choice
(e.g., the time component of the metric fluctuation is set zero 
in the synchronous gauge), {\it but} by a physical condition for the 
proper-time hypersurface that can be expressed in any gauge choice.

Beyond the linear order in perturbations,
a spatial gauge choice also plays an important role. The diffeomorphism
symmetry allows four degrees of freedom in coordinates that can be chosen
by a temporal gauge and a spatial gauge. A choice of temporal gauge
amounts to a choice of a hypersurface of simultaneity or a time-slicing,
and a choice of spatial gauge amounts to a choice of a spatial gridding
in a given hypersurface.  In the past, almost exclusively spatial C-gauge
has been used 
for a spatial gauge choice, in which there is no off-diagonal component of 
scalar and vector fluctuations in the spatial component of the metric tensor
(see \cite{YOO14b} for the terminology). In certainty this is not the only
choice one can make, and indeed various spatial gauge choices have been
explored in recent years in implementing $N$-body simulations in a
relativistic framework \cite{FIRAET15,FITRET16,FITRET17,ADFI19}
or developing a relativistic version of the Lagrangian perturbation theory
 \cite{RAMPF14,RAWI14}.
In addition to a spatial C-gauge choice, galaxy bias has been also modeled 
in spatial B-gauge \cite{BRHIET14,BRHIWA14,KOUMET18,UMKOET19,UMKO19}, 
in which there is no fluctuation in the space-time 
component of the metric tensor (see \cite{YOO14b} for the terminology). 
Other choices of spatial gauge are also allowed in general relativity. 

Now consider a general coordinate transformation 
\beeq
\label{ct}
\tilde x^\mu=e^{\zeta^\nu\pa_\nu}x^\mu=x^\mu+\zeta^\mu+\frac12\zeta^\mu{}_{,\nu}
\zeta^\nu+\OO(3)~,
\eneq
parametrized by an infinitesimal vector field~$\zeta^\mu$.
Any perturbation quantities~$\de\bdv{T}$
gauge transform \cite{STWA74,BRMAET97,YODU17}
as
\beeq
\label{LIE}
\widetilde{\de \bdv{T}}
=\de \bdv{T}-\pounds_\zeta\bar {\bdv{T}}+\frac12\pounds_\zeta\pounds_\zeta
\bar {\bdv{T}}-\pounds_\zeta\de \bdv{T}+\OO(3)~,
\eneq
where we again split a field~$\bdv{T}:=\bar{\bdv{T}}+\de \bdv{T}$ into
a background~$\bar{\bdv{T}}$ and a perturbation~$\de\bdv{T}$ and 
we suppressed its tensorial indicies. At the linear order in perturbations, 
only the Lie derivative $\pounds_\zeta\bar{\bdv{T}}$ contributes
in the gauge transformation, and it vanishes
for a spatial component of~$\zeta^\mu$, if~$\bar{\bdv{T}}$ is a scalar quantity.
In other words, due to the homogeneity and isotropy of the background universe,
any background quantities~$\bar{\bdv{T}}$  are just a function 
of time only, and hence a choice of spatial gauge condition 
is {\it irrelevant} at the linear-order calculations of scalar quantities.
Beyond the linear order in perturbations, {\it a spatial gauge choice} makes
a difference in gauge transformation (this is true even at the linear order,
if~$\bdv{T}$ is a tensor).

To be specific, we show that
the ambiguities in spatial gauge choice beyond the linear
order in perturbations have a direct impact on the three-point correlation
function or the bispectrum, whose leading order contribution arises from
the second-order contribution in the field. Consider a spatial gauge
transformation~$\LL^i$ in a hypersurface fixed by a choice of temporal gauge 
($T\equiv0$), and the matter density fluctuation (or any scalar quantities)
then gauge transform as
\beeq
\label{GT2}
\tilde\delta_m(x^\mu)=\delta_m(x^\mu)-\LL^i~\nabla_i\delta_m+\OO(3)~,
\eneq
where the time coordinates in~$x^\mu$ and~$\tilde x^\mu$ are identical.
It is readily evident that the three-point correlation function
(or the bispectrum) also gauge transforms with a spatial gauge choice,
even when a temporal gauge condition is completely fixed. 

For example,
a single-field consistency relation in inflationary models is obtained
by computing the bispectrum of a comoving-gauge curvature 
perturbation~$\cal R$ in the squeezed triangular configuration 
\cite{MALDA03,CRZA04,CHHUET07,SEZA12}. 
In this limit, the bispectrum
is in proportion to $n_s-1$, as long as there exists only one scalar degree
of freedom, regardless of its dynamics, and hence an observational
test of this consistency relation would provide a powerful way of understanding
the early Universe, where $n_s$~is the spectral index of the
curvature perturbation. However, note that while~$\cal R$ is independent
of a spatial gauge choice at the linear order, it depends on a spatial
gauge choice beyond the linear order. In computing the consistency relation,
a choice of 
spatial C-gauge was exclusively adopted, though other choices of spatial gauge
are possible. It was shown in \cite{MIYO22a}
that this consistency relation in the squeezed bispectrum changes according
to a spatial gauge choice and hence can be made in proportion to
the running $\alpha_s$~of the spectral index, rather than~$n_s$ itself,
by an appropriate gauge choice. A key point in \cite{MIYO22a}
is that the consistency 
relation depends on a spatial gauge choice in addition to the temporal
gauge choice, because it
is {\it not} a direct observable, on which we further elaborate below.

In modeling galaxy bias, the question was ``if we were to fix a gauge choice 
by hand, which gauge condition do we have to choose and why?'' For a choice
of temporal gauge, the physical argument for a proper-time hypersurface
uniquely fixes a time-slicing that can be expressed in any gauge choice.
For a choice of spatial gauge, however, 
the physical argument for a proper-time hypersurface {\it does not}
fix a spatial gridding. It was
shown \cite{YOO14b} that while both the synchronous and the comoving gauges
correspond to a proper-time hypersurface, they differ by a spatial gauge
choice, and their matter density fluctuations are different at the second
order in perturbations, leaving ambiguities in the linear bias relation 
in eq.~\eqref{linear}. In summary, there is currently {\it no}
physical argument for modeling galaxy bias in literature
that uniquely fixes a spatial gridding and can be
expressed in any choice of gauge.

\subsection{Second-order bias and primordial non-Gaussianity}
Beyond the
linear order in perturbations, there exist additional contributions in 
the standard (Newtonian) galaxy bias model (see, e.g., 
\cite{SZALA88,FRGA93,FRY96,MCDON06,MCRO09,SENAT15,MISCZA15,ANFAET15,DEJESC18}).
Up to the second order,
two extra bias parameters are needed 
\beeq
\label{2nd}
\de_g=b_1\de_m+\frac12b_2\de_m^2+b_ss^2+\OO(3)
\eneq
where the traceless tidal tensor is
\beeq
s_{ij}:=\left(\Delta^{-1}\nabla_i\nabla_j-\frac13\de_{ij}\right)\de_m~,
\Dquad s^2:=s_{ij}s_{ij}~.
\eneq
Additional contributions are added (or prohibited)
at each order \cite{MCDON06,MCRO09}
to yield contributions to~$\de_g$ that is
compatible with the symmetry in the system, or the Galilean symmetry 
in Newtonian dynamics.
This is the essence of the effective field theory (EFT) approach
(see, e.g., \cite{SENAT15,MISCZA15,ANFAET15}),
and for instance a gravitational potential 
contribution~$\phi$ in eq.~\eqref{2nd} is not allowed as a constant shift
in~$\phi$ everywhere {\it cannot}
 be measured according to the equivalence principle.

As discussed in the introduction, this standard modeling of galaxy bias 
works well on small scales, but it is {\it incomplete} in general
relativity. First, a spatial gauge transformation would yield 
\beeq
\tilde\de_g(x)=b_1\de_m+\frac12b_2\de_m^2+b_ss^2-b_1{\cal L}
\cdot\nabla\de_m+\OO(3)~,
\eneq
with an extra contribution arising, from the transformation of the
matter fluctuation in eq.~\eqref{GT2}. The spatial 
transformation~${\cal L}^i$ is {\it fully arbitrary}, allowed by diffeomorphism
symmetry. Hence this term cannot be absorbed
into the existing bias parameters,
signaling the breakdown of the EFT modeling
of galaxy bias in Newtonian dynamics
from the perspective of relativistic dynamics. However, mind that 
there exists a case, in which this is possible,
 as the Newtonian limit is part of general relativity. 
For a specific transformation from the Lagrangian frame to the Eulerian
frame in the Newtonian dynamics, the spatial transformation~${\cal L}^i$
corresponds to the displacement field~$\Psi^i$, i.e.,  
${\cal L}^i\equiv\Psi^i
=-\Delta^{-1}\nabla^i\de_m$, which can be further decomposed in terms 
of~$\de_m^2$, $s^2$, and the second-order matter density~$\de_m^{(2)}$,
such that the extra term can be absorbed into the existing bias parameters.
In general, however, we emphasize again that the spatial transformation
is fully arbitrary, not limited to this specific case.

Furthermore, the recent analysis of the relativistic effects
reveals \cite{YOFIZA09,YOO10,BODU11,CHLE11,JESCHI12,YOO14a}
that there exist numerous relativistic effects in galaxy clustering 
such as the gravitational potential~$\phi$. Such contributions are indeed
allowed, given the diffeomorphism symmetry in general relativity.
In fact, compatibility  with the equivalence
principle, or the independence of a constant shift in~$\phi$ is enforced
in galaxy clustering
\cite{JESCHI12,BIYO17,SCYOBI18,GRSCET20,BAYO21},
{\it not by prohibiting} such contributions, {\it but by allowing} 
other relativistic contributions of~$\phi$
that cancel the effect of a constant shift in~$\phi$.
It is evident that an EFT approach to modeling galaxy bias in relativistic
dynamics is more involved than the standard EFT approaches in Newtonian
dynamics.
This incompatibility in the standard model for galaxy bias 
comes from the fact that the diffeomorphism symmetry
in general relativity is {\it not} respected in the standard model
for galaxy bias, or
there is {\it no} covariant description of galaxy bias.

We stress again that Newtonian dynamics and the standard model for galaxy  
bias, though incompatible with general relativity,
work well in most cases of our cosmological interest. As discussed,
however, the relativistic effects become important, when the signals of
our cosmological interest are also small relativistic effects such as
the primordial non-Gaussianity.
The deviation from a perfect Gaussianity in the primordial fluctuations
contains crucial clues about the early Universe. The primordial
non-Gaussianity is often parametrized \cite{KOSP01,BAKOET04} as
\beeq
\label{FNL}
{\cal R}(x)={\cal R}_g(x)+\frac35\fnl{\cal R}_g^2(x)~,
\eneq
a local-type deviation from a Gaussian curvature fluctuation~${\cal R}_g$ 
in the comoving gauge, and the prediction for the standard single-field 
inflationary model is of the order $\fnl\sim(n_s-1)$ \cite{MALDA03},
while other non-standard inflationary models often predict
larger non-Gaussianities $\fnl\geq1$ \cite{SELI05a,BYSAWA06,CHHUET07,DETS11}.
Constraining the level of primordial non-Gaussianity in our Universe
is one of the key targets in large-scale
surveys such as the Vera C.~Rubin Observatory \cite{LSST04}, 
the Euclid mission \cite{EUCLID11},
and the Nancy Grace Roman Space Telescope \cite{WFIRST12}.

In the presence of a local-type primordial non-Gaussianity, it was shown
\cite{DADOET08,SLHIET08} that the local rms fluctuations are modulated,
affecting the probability of the matter density fluctuation above
the threshold $\de_c\simeq1.686$, and hence  the galaxy number density
\beeq
\label{DALAL}
\de_g=b~\de_m+2\fnl(b-1)\de_c\phi_p+\OO(2)~,
\eneq
receives additional contribution from the primordial gravitational potential 
$\phi_p:=(3/5){\cal R}$. Compared to the first term~$b~\de_m$, 
this extra contribution makes the bias parameter scale dependent
with respect to the matter density,
as $\phi_p\propto\de_m/k^2$, boosting the galaxy power spectrum on large scales,
where the bias parameter is believed to be a scale-independent constant.
Measurements of galaxy clustering in the upcoming surveys 
would then provide a clear path to best constraining the primordial 
non-Gaussianity, given that more large-scale modes are available for
galaxy surveys than in CMB observations (see, e.g.,
\cite{LSST04,EUCLID11,WFIRST12,YOHAET12}).

However, at first glance, there exist a few unsatisfactory features:
the galaxy number density in eq.~\eqref{DALAL} responds to a
shift in the uniform gravitational potential, and the second-order
correction in eq.~\eqref{FNL} appears in eq.~\eqref{DALAL} as the first-order
correction. A more rigorous derivation for galaxy clustering
was made \cite{MAVE08} in Newtonian dynamics
by using the spherical
collapse model in the presence of primordial non-Gaussianity \cite{MALUBO86},
and the two-point correlation function~$\xi_g$ for galaxy,
\beeq
\label{MATA}
\xi_g(x_1,x_2)={\nu^2\over\sigma^2_R}~\xi_m(x_1,x_2)
+{\nu^3\over\sigma^3_R}~\xi_m(x_1,x_1,x_2)~,
\eneq
receives an additional contribution from the three-point matter correlation
function~$\xi_m$ in the squeezed triangular configuration,
where $\sigma_R$ is the rms matter fluctuation smoothed by a scale~$R$,
$\nu:=\de_c/\sigma_R$ is the scaled threshold for the spherical 
collapse model, and the bias parameter is then $b=1+\nu/\sigma_R$
in this model \cite{KAISE84,POWI84,BBKS86,BCEK91}.

Indeed, the non-Gaussian
correction in eq.~\eqref{MATA} is the next order in perturbations,
compared to the leading two-point correlation function,
as indicated in eq.~\eqref{FNL} with a non-vanishing~$\fnl$. 
While things are in order in Newtonian dynamics, this implies
that the ambiguities in spatial gauge due to the transformation in
eq.~\eqref{GT2} affect the correction from the three-point correlation
function in eq.~\eqref{MATA} in relativistic dynamics. 
Such relativistic effects may affect the measurements of~$\fnl$, 
at the level of order unity
\cite{CAMASA15,BRHIET14,BRHIWA14,VIVEMA14,MAPIRO21}. Note that the
prediction for~$\fnl$ in the standard inflationary model is negligibly
small \cite{MALDA03},
such that $\Delta\fnl\sim$few from the relativistic effects in general
relativity might be misinterpreted as the signal from inflationary models
beyond the standard model. Given the enormous
potential to probe the early Universe, the theoretical description 
in eqs.~\eqref{DALAL} or~\eqref{MATA} needs to be further improved 
in a relativistic framework for an accurate prediction for~$\Delta\fnl$
from the relativistic effects.

\subsection{Spatial gauge choices for galaxy bias in literature} 
The linear bias relation
in eq.~\eqref{linear} is expected to be valid with the matter density
fluctuation beyond the linear order in perturbations, and we need a choice
of spatial gauge in addition to a choice of temporal gauge, if we want
to impose the linear bias relation in a specific gauge.
Given that a temporal gauge choice for a proper-time hypersurface is physically 
justified and also universally adopted in literature,
we discuss the past choices for spatial gauge and aruge that 
there is {\it no} good physical argument
for a specific choice of spatial gauge so far in literature. 
The  diffeomorphism symmetry in general relativity puts all choices 
on a equivalent footing.

The temporal comoving gauge ($v\equiv0$)
with spatial C-gauge was preferred in \cite{YOO14b} (or gauge-I therein), 
since the equation of motion for a pressureless medium in this gauge
 is identical to one in Newtonian dynamics. However, this correspondence
is valid only up to the second order in perturbations 
\cite{HWNO05,HWNO05b,HWNO07b}, or {\it not} valid beyond the second order
in perturbations. Furthermore,
the solution in this gauge contains {\it not only} the Newtonian contributions,
{\it but also} the relativistic contributions 
\cite{BRHIET14,BRHIWA14,VIVEMA14,BEBAET15,YOGO16}, which
weakens the arguments for this gauge choice. Furthermore, 
following the argument
in \cite{PEEBL80}, this gauge choice was advocated in \cite{YOO14b}, as
it preserves the ``mass conservation,'' which in fact meant that the
leading terms in the spatial expansion of~$\de_m$ add up to zero. 
The ``momentum conservation'' was used in \cite{PEEBL80}
to refer to the vanishing sum of
the next-leading terms in the spatial expansion. 
Unfortunately, this argument for the spatial C-gauge used in \cite{YOO14b}
is just a statement about the coordinate properties, rather than a physical
argument for supporting the gauge choice. Moreover, with
the relativistic effects properly considered in this gauge
\cite{BRHIET14,BRHIWA14,VIVEMA14,BEBAET15,YOGO16},
the condition for the vanishing sum in the leading terms is in fact {\it not}
satisfied, i.e., the ``mass conservation'' argument
used in \cite{YOO14b} is {\it invalid}. Furthermore,
the synchronous gauge ($\alpha\equiv0$ with spatial B-gauge
or gauge-II in \cite{YOO14b}) was shown to correspond to one 
in the Lagrangian frame (see also \cite{RAMPF14,RAWI14}),
such that this coordinate is also related by a gauge transformation
to the temporal comoving gauge with spatial C-gauge. Similar to
the Newtonian dynamics, these two gauge choices reflect two different
views of the same system with {\it no} preference to one or the other.

Another approach in literature that gained popularity is the conformal
Fermi coordinate (CFC; 
\cite{PASCZA13,DAPASC15,DAPASC15a,CAPASC17,IPSC17,UMKOET19}),
in which the local metric along a given
geodesic parametrized by a proper time~$\tau$
is $g_{\mu\nu}=a^2(\tau)\eta_{\mu\nu}$ and the corrections to the
metric in the neighborhood appear at least quadratically in distances.
It provides a useful framework to interpret Newtonian calculations in
general relativity, as it closely matches Newtonian coordinates in the
neighborhood. 
However, CFC is {\it not} Newtonian either, simply because the metric
around the world line deviates from the Minkowski metric~$\eta_{\mu\nu}$
(or Euclidean even). 
Numerical simulations can be used to study galaxy bias with the relativistic
effects accounted for \cite{FIRAET15,FITRET16,ADDAET16,ADDAET16b,BOBEPO17}. In 
particular, the light-cone observations were made \cite{BOBEPO17,LESCET23}
in the simulations, and the galaxy bias parameters obtained in the light-cone
observations were compared to the bias parameters on a hypersurface in 
the simulations. While these approaches provide a powerful tool to study
the relativistic effects and the light-cone observables, the focus of 
this work is galaxy bias in a relativistic framework, or
the relativistic relation between the galaxy and the matter
density distributions, one step before we account for the light propagation.
In both cases, the spatial C-gauge was chosen in the
numerical simulations.

In short, Newtonian correspondence in relativistic dynamics is a good
consistency check, but it is {\it not} a sufficient condition for validating
a model in general relativity. In the end, general relativity is {\it not} 
Newtonian. Figure~\ref{matter} highlights this point
that most of 
the matter fluctuations in various gauges correspond to the Newtonian
matter fluctuation on small scales, while they are different on large scales.
We have found {\it no} good physical justification
for any specific choice of spatial gauge for galaxy bias. 
 However, as we show below, a covariant description of galaxy
bias resolves the issue of choosing a gauge condition in a way that
any gauge choice is on a equal footing, consistent with the diffeomorphism
symmetry.

\section{Gauge-invariance of cosmological observables} 
\subsection{Cosmological observables}
\label{sec:obs}
As opposed to the
coordinate dependence discussed so far, the observers utilize a {\it unique}
coordinate system (up to a trivial rotation)
to map the cosmological sources in the rest frame, i.e.,
observed angular positions~$\Vang$ and observed redshifts~$z$. 
These numbers and the observer coordinates 
are independent of coordinates we use in our theoretical description 
of the Universe, i.e., FRW coordinates. Indeed, the mapping of the observers
is {\it not} a coordinate, as it does {\it not} cover the whole spacetime 
manifold, nor is it invertible (see \cite{MIYO20} for further discussion).
In terms of galaxy clustering, the observers simply count the number of
the observed galaxies in a small volume defined by a small
redshift bin and a small solid angle in observations, and this number
yields the observed galaxy number density as a function of the observed
position $(z,n^i)$ \cite{YOFIZA09}.
Therefore, it is evident that our theoretical description for
the observed galaxy
number density fluctuation should be a scalar invariant under diffeomorphism
\beeq
\label{inv}
\de^\up{obs}_g(x^\mu_s)=\tilde{\de}^\up{obs}_g(\tilde x^\mu_s)~,
\eneq
or gauge-invariant when expressed in terms of the observed position
\cite{YODU17} to properly compare to observations.
This is indeed the case 
\cite{YOZA14,DIDUET14,BEMACL14,BEMACL14b,DIDUET16,UMJOET17,JOUMET17,YODU17,MIYO20,MAYO22}
as the source galaxy position in a given coordinate is
\beeq
x^\mu_s=:\bar x^\mu_z+\Delta x^\mu~,
\eneq
where the observed position of the source specified by~$(z,n^i)$
in a coordinate is expressed as~$\bar x^\mu_z$
and its deviation from the real position~$x^\mu_s$ in a coordinate
is captured by~$\Delta x^\mu$.
For instance,
the expression up to the second order in perturbations is
\beeq
\de_g^\up{obs}(x_s^\mu)=\de^\up{obs}_g(\bar x_z^\mu)
+\Delta x^\mu\pa_\mu\de_g^\up{obs}+\OO(3)~.
\eneq
With the observed position~$\bar x^\mu_z$ fixed as a reference in all
coordinates, 
the deviation~$\Delta x^\mu$ gauge transforms, according to the coordinate
transformation in eq.~\eqref{ct}, and this transformation cancels
any temporal and spatial transformation of~$\de_g^\up{obs}(x^\mu_s)$
in eq.~\eqref{LIE} [or Eqs.~\eqref{galaxy} and~\eqref{GT2}],
rendering the observed galaxy number density fluctuation
{\it fully gauge invariant at the observed position}.

The theoretical expression for the observed galaxy number density
fluctuation~$\de_g^\up{obs}$ is derived 
by solving the geodesic equation
and relating the observed position~$\bar x^\mu_z$ to the source position~$x^\mu$
in a given coordinate, which uniquely determines the volume effect \cite{YOO09}
such as the gravitational lensing \cite{KAISE92}
and the redshift-space distortion \cite{KAISE87}.
However, being a key part of the source effect, the galaxy bias part
in the theoretical description remains ambiguous beyond the linear order
in perturbations due to the lack of covariant galaxy bias model.

\subsection{A covariant toy model for galaxy bias} 
It is in fact simple to devise
a {\it covariant} model for galaxy bias, while the difficult part
lies in building a {\it realistic} model. For instance, 
the simplest one is to model the
galaxy distribution as the matter distribution 
or a function of the matter distribution:
\beeq
n_g(x^\mu)\equiv {\cal N}\rho_m^b(x^\mu)~,
\eneq
where we assumed a power-law function for simplicity and $\cal N$~is a 
dimensionful constant.
This simple model is covariant, and the bias relation in eq.~\eqref{linear} 
is valid in all coordinates, with~$b$ in the power-law as the linear
bias parameter:
\beeq
n_g(x^\mu)={\cal N}\bar\rho_m^b(t)\left[1+b~\de_m(x)+\cdots\right]~.
\eneq
The evolution of the galaxy number density in this model is, however, 
related to the bias parameter~$b$, according to eq.~\eqref{evol}. Hence,
the model is {\it not} observationally viable, which illustrates an
extra complexity in developing a covariant model for galaxy bias ---
one has to account for the time evolution of the galaxy number density 
in building a model for galaxy bias. Note that the time evolution of galaxy
bias can be accounted for in the standard approaches (e.g., \cite{FRY96}), 
but 
this aspect is {\it not} a necessary condition in the standard model.

To implement the physical arguments for the proper-time hypersurface
into the model
and to match the observational constraint on the evolution of the galaxy
number density, we develop another simple {\it toy} model by 
considering an extra function~$E(t_p)$ 
of proper-time~$t_p$ of the galaxy along the world line:
\beeq
\label{cova}
n_g(x^\mu)\equiv {\cal N}\rho_m^b(x^\mu)E(t_p)=:{\cal N}
\rho_m^b(x^\mu)(1+z_p)^{b_t}~,
\eneq
where we again assumed a simple power-law function for~$E(t_p)$ and expressed
it in terms of the redshift parameter~$z_p$ in the proper-time hypersurface,
 and $\cal N$~is another dimensionful
constant of the model. The functional dependence of the model on~$t_p$
prefers the proper-time hypersurface, and the evolution of the galaxy
number density is 
\beeq
{d\ln\bar n_g\over d\ln(1+z)}=3b+b_t~, 
\eneq
instead of~$3b$, which allows freedom to match the observed evolution.
The galaxy number density can be expanded in the proper-time hypersurface as
\beeq
n_g(x^\mu)={\cal N}\bar\rho_m^b(t_p)(1+z_p)^{b_t}
\left[1+b~\de_p(x)+\cdots\right]~,
\eneq
and a choice of spatial gauge for galaxy bias is {\it not} needed, as it is
covariant in this model, where $\de_p$ is the matter density fluctuation
in the proper-time hypersurface.
To illustrate the difference, the question in the standard approaches is
to use the matter density fluctuation with which gauge conditions:
\beeq
\de_g(x)=b~\de_m^\up{I}(x)~? \Dquad \de_g(x)=b~\de_m^\up{II}(x)~?
\Dquad \de_g(x)=\cdots ?
\eneq
where the super-scripts~I and~II indicate the matter density fluctuation
computed in various choices of gauge conditions, either temporal or spatial.
Note that the matter density fluctuation~$\de_m(x)$ is uniquely
defined in each gauge and they are all different.  According to the
covariant toy model, the galaxy bias relation in contrast becomes
\beeq
\de_g^\up{I}(x^\up{I})=b~\de_p^\up{I}(x^\up{I})~,\Dquad
\de_g^\up{II}(x^\up{II})=b~\de_p^\up{II}(x^\up{II}),~\Dquad
\de_g^\up{I}(x^\up{I})=\de_g^\up{II}(x^\up{II})=\cdots~,
\eneq
such that the same bias factor is multiplied by the matter density fluctuation
in each gauge choice and the galaxy fluctuation at a given physical point 
({\it not} a coordinate position, $x^\up{I}\neq x^\up{II}$) 
is identical in any gauge choice. 
Therefore, any spatial gauge choice is allowed with the same bias parameter~$b$
and no physical justification for a specific choice of spatial gauge 
is needed, resolving the issues associated with a spatial gauge choice.

Consequently, when the galaxy two-point correlation~$\langle\de_g\de_g\rangle$
(or for that matter the $N$-point correlation as well)
is computed, the outcome will differ, depending on which gauge choice is
adopted, as shown in Figure~\ref{matter}.
However, as discussed in Section~\ref{sec:obs} or shown in 
eq.~\eqref{inv}, the outcome will be identical and independent of gauge choice,
when expressed in terms of
the observed position for a covariant model for galaxy bias.

While these toy models serve the purpose of demonstrating the need to develop
a covariant model, they are {\it not} realistic models for galaxy bias.
A realistic covariant model for galaxy bias can be developed by generalizing,
for example, the peak model \cite{BBKS86,BOMY96,DESJA13}.
As discussed, a temporal
gauge choice to the proper-time hypersurface can be enforced {\it not} 
by choosing a coordinate by hand, {\it but} by adopting a four velocity~$u^\mu$ 
of the galaxy sample as the time direction in the model. The matter
overdensity in the proper-time hypersurface should then satisfy the condition
for a collapse~$\de_m\geq\de_c$, and the other condition that the peak position
should have a vanishing spatial derivative and a negative second spatial
derivative can be accommodated in a covariant manner by defining 
the spatial projection tensor~$\HH_{\mu\nu}:=g_{\mu\nu}+u_\mu u_\nu$
and by using the projected covariant derivative 
$\nabla_\mu^{(3)}:=\HH^\nu_\mu\nabla_\nu$ as a
spatial derivative in the proper-time hypersurface, where $g_{\mu\nu}$ is
the metric tensor. In this way it would be rather straightforward to build
a covariant peak model, but of course quantifying the impact of the covariant
peak model compared to the standard model 
requires a further investigation, beyond the scope of the current manuscript.

\section{Conclusions}
The standard model for galaxy bias (and many
other elements in cosmology as well) is built on a perturbative framework 
in Newtonian cosmology, and hence it lacks a covariant expression 
for galaxy bias. We have demonstrated that the reverse-engineered approaches
to modeling galaxy bias in literature
{\it do not} respect the diffeomorphism symmetry in general
relativity. 

To mitigate this issue,
we have considered a simple, but covariant toy model in eq.~\eqref{cova}
that respects the diffeomorphism symmetry in general relativity
and hence resolves all the ambiguities in the standard model for galaxy bias.
One obvious deficiency in the toy model is that galaxies can form even in the
background universe --- It is the matter fluctuations in the local neighborhood,
not the absolute value of the matter density that matters for galaxy formation;
in the early Universe, galaxies or even dark matter halos cannot form,
despite the enormous matter density everywhere.
Though this consideration can be put into
the model by adjusting the function~$E(t_p)$ in eq.~\eqref{cova}, 
a more physical model for galaxy bias remains to be built. In particular, 
building a realistic model for galaxy bias, especially in the presence
of the primordial non-Gaussianity, deserves a further in-depth investigation,
and a possibility is to put the peak model for galaxy bias in a relativistic 
framework.

Though strictly true in principle,  the flaws in the standard
approaches pose {\it no} significant issues in practice,  as the 
applications of galaxy clustering in most cases are well described by
the Newtonian dynamics. This work explores the subtle issues in a regime,
where the relativistic effects become important. Given the level 
of the measurement uncertainties~$\Delta\fnl\approx$ few
in measuring the primordial non-Gaussianity in the upcoming large-scale
surveys (e.g., \cite{LSST04,EUCLID11,WFIRST12,ANARET18,MUREET22,CAIVET22,DALEET22}),
a covariant relativistic description of galaxy bias will play
a crucial role in extracting such small relativistic effects that 
contain yet important cosmological information
\cite{YOGRMI22,CADI22,FOPAET23}.

\acknowledgments

We acknowledge useful discussions with Ermis Mitsou and Julian Adamek.
This work is supported by the Swiss National Science Foundation
and a Consolidator Grant of the European Research Council.

\bibliography{bias.bbl}

\begin{thebibliography}{118}
\expandafter\ifx\csname natexlab\endcsname\relax\def\natexlab#1{#1}\fi
\expandafter\ifx\csname bibnamefont\endcsname\relax
  \def\bibnamefont#1{#1}\fi
\expandafter\ifx\csname bibfnamefont\endcsname\relax
  \def\bibfnamefont#1{#1}\fi
\expandafter\ifx\csname citenamefont\endcsname\relax
  \def\citenamefont#1{#1}\fi
\expandafter\ifx\csname url\endcsname\relax
  \def\url#1{\texttt{#1}}\fi
\expandafter\ifx\csname urlprefix\endcsname\relax\def\urlprefix{URL }\fi
\providecommand{\bibinfo}[2]{#2}
\providecommand{\eprint}[2][]{\url{#2}}

\bibitem[{\citenamefont{{Press} and {Schechter}}(1974)}]{PRSC74}
\bibinfo{author}{\bibfnamefont{W.~H.} \bibnamefont{{Press}}} \bibnamefont{and}
  \bibinfo{author}{\bibfnamefont{P.}~\bibnamefont{{Schechter}}},
  \bibinfo{journal}{\apj} \textbf{\bibinfo{volume}{187}}, \bibinfo{pages}{425}
  (\bibinfo{year}{1974}).

\bibitem[{\citenamefont{{Kaiser}}(1984)}]{KAISE84}
\bibinfo{author}{\bibfnamefont{N.}~\bibnamefont{{Kaiser}}},
  \bibinfo{journal}{\apjl} \textbf{\bibinfo{volume}{284}}, \bibinfo{pages}{L9}
  (\bibinfo{year}{1984}).

\bibitem[{\citenamefont{{Politzer} and {Wise}}(1984)}]{POWI84}
\bibinfo{author}{\bibfnamefont{H.~D.} \bibnamefont{{Politzer}}}
  \bibnamefont{and} \bibinfo{author}{\bibfnamefont{M.~B.}
  \bibnamefont{{Wise}}}, \bibinfo{journal}{\apjl}
  \textbf{\bibinfo{volume}{285}}, \bibinfo{pages}{L1} (\bibinfo{year}{1984}).

\bibitem[{\citenamefont{{Bardeen} et~al.}(1986)\citenamefont{{Bardeen}, {Bond},
  {Kaiser}, and {Szalay}}}]{BBKS86}
\bibinfo{author}{\bibfnamefont{J.~M.} \bibnamefont{{Bardeen}}},
  \bibinfo{author}{\bibfnamefont{J.~R.} \bibnamefont{{Bond}}},
  \bibinfo{author}{\bibfnamefont{N.}~\bibnamefont{{Kaiser}}}, \bibnamefont{and}
  \bibinfo{author}{\bibfnamefont{A.~S.} \bibnamefont{{Szalay}}},
  \bibinfo{journal}{\apj} \textbf{\bibinfo{volume}{304}}, \bibinfo{pages}{15}
  (\bibinfo{year}{1986}).

\bibitem[{\citenamefont{{Bond} et~al.}(1991)\citenamefont{{Bond}, {Cole},
  {Efstathiou}, and {Kaiser}}}]{BCEK91}
\bibinfo{author}{\bibfnamefont{J.~R.} \bibnamefont{{Bond}}},
  \bibinfo{author}{\bibfnamefont{S.}~\bibnamefont{{Cole}}},
  \bibinfo{author}{\bibfnamefont{G.}~\bibnamefont{{Efstathiou}}},
  \bibnamefont{and} \bibinfo{author}{\bibfnamefont{N.}~\bibnamefont{{Kaiser}}},
  \bibinfo{journal}{\apj} \textbf{\bibinfo{volume}{379}}, \bibinfo{pages}{440}
  (\bibinfo{year}{1991}).

\bibitem[{\citenamefont{{Desjacques} et~al.}(2018)\citenamefont{{Desjacques},
  {Jeong}, and {Schmidt}}}]{DEJESC18}
\bibinfo{author}{\bibfnamefont{V.}~\bibnamefont{{Desjacques}}},
  \bibinfo{author}{\bibfnamefont{D.}~\bibnamefont{{Jeong}}}, \bibnamefont{and}
  \bibinfo{author}{\bibfnamefont{F.}~\bibnamefont{{Schmidt}}},
  \bibinfo{journal}{\physrep} \textbf{\bibinfo{volume}{733}},
  \bibinfo{pages}{1} (\bibinfo{year}{2018}), \eprint{1611.09787}.

\bibitem[{\citenamefont{{Eisenstein} et~al.}(2005)\citenamefont{{Eisenstein},
  {Zehavi}, {Hogg}, {Scoccimarro} et~al.}}]{EIZEET05}
\bibinfo{author}{\bibfnamefont{D.~J.} \bibnamefont{{Eisenstein}}},
  \bibinfo{author}{\bibfnamefont{I.}~\bibnamefont{{Zehavi}}},
  \bibinfo{author}{\bibfnamefont{D.~W.} \bibnamefont{{Hogg}}},
  \bibinfo{author}{\bibfnamefont{R.}~\bibnamefont{{Scoccimarro}}},
  \bibnamefont{et~al.}, \bibinfo{journal}{\apj} \textbf{\bibinfo{volume}{633}},
  \bibinfo{pages}{560} (\bibinfo{year}{2005}), \eprint{arXiv:astro-ph/0501171}.

\bibitem[{\citenamefont{{Tegmark} et~al.}(2006)}]{TEEIET06}
\bibinfo{author}{\bibfnamefont{M.}~\bibnamefont{{Tegmark}}}
  \bibnamefont{et~al.}, \bibinfo{journal}{\prd} \textbf{\bibinfo{volume}{74}},
  \bibinfo{pages}{123507} (\bibinfo{year}{2006}),
  \eprint{arXiv:astro-ph/0608632}.

\bibitem[{\citenamefont{{Anderson} et~al.}(2012)\citenamefont{{Anderson},
  {Aubourg}, {Bailey}, {Bizyaev}, {Blanton}, {Bolton}, {Brinkmann},
  {Brownstein} et~al.}}]{ANAUET12}
\bibinfo{author}{\bibfnamefont{L.}~\bibnamefont{{Anderson}}},
  \bibinfo{author}{\bibfnamefont{E.}~\bibnamefont{{Aubourg}}},
  \bibinfo{author}{\bibfnamefont{S.}~\bibnamefont{{Bailey}}},
  \bibinfo{author}{\bibfnamefont{D.}~\bibnamefont{{Bizyaev}}},
  \bibinfo{author}{\bibfnamefont{M.}~\bibnamefont{{Blanton}}},
  \bibinfo{author}{\bibfnamefont{A.~S.} \bibnamefont{{Bolton}}},
  \bibinfo{author}{\bibfnamefont{J.}~\bibnamefont{{Brinkmann}}},
  \bibinfo{author}{\bibfnamefont{J.~R.} \bibnamefont{{Brownstein}}},
  \bibnamefont{et~al.}, \bibinfo{journal}{\mnras}
  \textbf{\bibinfo{volume}{427}}, \bibinfo{pages}{3435} (\bibinfo{year}{2012}),
  \eprint{1203.6594}.

\bibitem[{\citenamefont{{Aubourg} et~al.}(2015)\citenamefont{{Aubourg},
  {Bailey}, {Bautista}, {Beutler} et~al.}}]{AUBAET15}
\bibinfo{author}{\bibfnamefont{{\'E}.}~\bibnamefont{{Aubourg}}},
  \bibinfo{author}{\bibfnamefont{S.}~\bibnamefont{{Bailey}}},
  \bibinfo{author}{\bibfnamefont{J.~E.} \bibnamefont{{Bautista}}},
  \bibinfo{author}{\bibfnamefont{F.}~\bibnamefont{{Beutler}}},
  \bibnamefont{et~al.}, \bibinfo{journal}{\prd} \textbf{\bibinfo{volume}{92}},
  \bibinfo{eid}{123516} (\bibinfo{year}{2015}), \eprint{1411.1074}.

\bibitem[{\citenamefont{{Bond} and {Myers}}(1996)}]{BOMY96}
\bibinfo{author}{\bibfnamefont{J.~R.} \bibnamefont{{Bond}}} \bibnamefont{and}
  \bibinfo{author}{\bibfnamefont{S.~T.} \bibnamefont{{Myers}}},
  \bibinfo{journal}{\apjs} \textbf{\bibinfo{volume}{103}}, \bibinfo{pages}{1}
  (\bibinfo{year}{1996}).

\bibitem[{\citenamefont{{Mo} and {White}}(1996)}]{MOWH96}
\bibinfo{author}{\bibfnamefont{H.~J.} \bibnamefont{{Mo}}} \bibnamefont{and}
  \bibinfo{author}{\bibfnamefont{S.~D.~M.} \bibnamefont{{White}}},
  \bibinfo{journal}{\mnras} \textbf{\bibinfo{volume}{282}},
  \bibinfo{pages}{347} (\bibinfo{year}{1996}), \eprint{arXiv:9512127}.

\bibitem[{\citenamefont{{Matsubara}}(1999)}]{MATSU99}
\bibinfo{author}{\bibfnamefont{T.}~\bibnamefont{{Matsubara}}},
  \bibinfo{journal}{\apj} \textbf{\bibinfo{volume}{525}}, \bibinfo{pages}{543}
  (\bibinfo{year}{1999}), \eprint{arXiv:9906029}.

\bibitem[{\citenamefont{{Catelan} et~al.}(2000)\citenamefont{{Catelan},
  {Porciani}, and {Kamionkowski}}}]{CAPOKA00}
\bibinfo{author}{\bibfnamefont{P.}~\bibnamefont{{Catelan}}},
  \bibinfo{author}{\bibfnamefont{C.}~\bibnamefont{{Porciani}}},
  \bibnamefont{and}
  \bibinfo{author}{\bibfnamefont{M.}~\bibnamefont{{Kamionkowski}}},
  \bibinfo{journal}{\mnras} \textbf{\bibinfo{volume}{318}},
  \bibinfo{pages}{L39} (\bibinfo{year}{2000}), \eprint{astro-ph/0005544}.

\bibitem[{\citenamefont{{Ma} and {Fry}}(2000)}]{MAFR00}
\bibinfo{author}{\bibfnamefont{C.-P.} \bibnamefont{{Ma}}} \bibnamefont{and}
  \bibinfo{author}{\bibfnamefont{J.~N.} \bibnamefont{{Fry}}},
  \bibinfo{journal}{\apj} \textbf{\bibinfo{volume}{543}}, \bibinfo{pages}{503}
  (\bibinfo{year}{2000}), \eprint{astro-ph/0003343}.

\bibitem[{\citenamefont{{Seljak}}(2000)}]{SELJA00}
\bibinfo{author}{\bibfnamefont{U.}~\bibnamefont{{Seljak}}},
  \bibinfo{journal}{\mnras} \textbf{\bibinfo{volume}{318}},
  \bibinfo{pages}{203} (\bibinfo{year}{2000}), \eprint{astro-ph/0001493}.

\bibitem[{\citenamefont{{Sheth} et~al.}(2001)\citenamefont{{Sheth}, {Mo}, and
  {Tormen}}}]{SMT01}
\bibinfo{author}{\bibfnamefont{R.~K.} \bibnamefont{{Sheth}}},
  \bibinfo{author}{\bibfnamefont{H.~J.} \bibnamefont{{Mo}}}, \bibnamefont{and}
  \bibinfo{author}{\bibfnamefont{G.}~\bibnamefont{{Tormen}}},
  \bibinfo{journal}{\mnras} \textbf{\bibinfo{volume}{323}}, \bibinfo{pages}{1}
  (\bibinfo{year}{2001}), \eprint{astro-ph/9907024}.

\bibitem[{\citenamefont{{Scoccimarro} et~al.}(2001)\citenamefont{{Scoccimarro},
  {Sheth}, {Hui}, and {Jain}}}]{SCSHET01}
\bibinfo{author}{\bibfnamefont{R.}~\bibnamefont{{Scoccimarro}}},
  \bibinfo{author}{\bibfnamefont{R.~K.} \bibnamefont{{Sheth}}},
  \bibinfo{author}{\bibfnamefont{L.}~\bibnamefont{{Hui}}}, \bibnamefont{and}
  \bibinfo{author}{\bibfnamefont{B.}~\bibnamefont{{Jain}}},
  \bibinfo{journal}{\apj} \textbf{\bibinfo{volume}{546}}, \bibinfo{pages}{20}
  (\bibinfo{year}{2001}), \eprint{astro-ph/0006319}.

\bibitem[{\citenamefont{{Berlind} and {Weinberg}}(2002)}]{BEWE02}
\bibinfo{author}{\bibfnamefont{A.~A.} \bibnamefont{{Berlind}}}
  \bibnamefont{and} \bibinfo{author}{\bibfnamefont{D.~H.}
  \bibnamefont{{Weinberg}}}, \bibinfo{journal}{\apj}
  \textbf{\bibinfo{volume}{575}}, \bibinfo{pages}{587} (\bibinfo{year}{2002}),
  \eprint{astro-ph/0109001}.

\bibitem[{\citenamefont{{Cooray} and {Sheth}}(2002)}]{COSH02}
\bibinfo{author}{\bibfnamefont{A.}~\bibnamefont{{Cooray}}} \bibnamefont{and}
  \bibinfo{author}{\bibfnamefont{R.}~\bibnamefont{{Sheth}}},
  \bibinfo{journal}{\physrep} \textbf{\bibinfo{volume}{372}},
  \bibinfo{pages}{1} (\bibinfo{year}{2002}), \eprint{astro-ph/0206508}.

\bibitem[{\citenamefont{{McDonald}}(2006)}]{MCDON06}
\bibinfo{author}{\bibfnamefont{P.}~\bibnamefont{{McDonald}}},
  \bibinfo{journal}{\prd} \textbf{\bibinfo{volume}{74}},
  \bibinfo{pages}{103512} (\bibinfo{year}{2006}),
  \eprint{arXiv:astro-ph/0609413}.

\bibitem[{\citenamefont{{McDonald} and {Roy}}(2009)}]{MCRO09}
\bibinfo{author}{\bibfnamefont{P.}~\bibnamefont{{McDonald}}} \bibnamefont{and}
  \bibinfo{author}{\bibfnamefont{A.}~\bibnamefont{{Roy}}},
  \bibinfo{journal}{\jcap} \textbf{\bibinfo{volume}{8}}, \bibinfo{pages}{20}
  (\bibinfo{year}{2009}), \eprint{0902.0991}.

\bibitem[{\citenamefont{{Baldauf} et~al.}(2012)\citenamefont{{Baldauf},
  {Seljak}, {Desjacques}, and {McDonald}}}]{BASEET12}
\bibinfo{author}{\bibfnamefont{T.}~\bibnamefont{{Baldauf}}},
  \bibinfo{author}{\bibfnamefont{U.}~\bibnamefont{{Seljak}}},
  \bibinfo{author}{\bibfnamefont{V.}~\bibnamefont{{Desjacques}}},
  \bibnamefont{and}
  \bibinfo{author}{\bibfnamefont{P.}~\bibnamefont{{McDonald}}},
  \bibinfo{journal}{\prd} \textbf{\bibinfo{volume}{86}},
  \bibinfo{pages}{083540} (\bibinfo{year}{2012}), \eprint{1201.4827}.

\bibitem[{\citenamefont{{Desjacques}}(2013)}]{DESJA13}
\bibinfo{author}{\bibfnamefont{V.}~\bibnamefont{{Desjacques}}},
  \bibinfo{journal}{\prd} \textbf{\bibinfo{volume}{87}}, \bibinfo{eid}{043505}
  (\bibinfo{year}{2013}), \eprint{1211.4128}.

\bibitem[{\citenamefont{{Schmidt} et~al.}(2013)\citenamefont{{Schmidt},
  {Jeong}, and {Desjacques}}}]{SCJEDE13}
\bibinfo{author}{\bibfnamefont{F.}~\bibnamefont{{Schmidt}}},
  \bibinfo{author}{\bibfnamefont{D.}~\bibnamefont{{Jeong}}}, \bibnamefont{and}
  \bibinfo{author}{\bibfnamefont{V.}~\bibnamefont{{Desjacques}}},
  \bibinfo{journal}{\prd} \textbf{\bibinfo{volume}{88}},
  \bibinfo{pages}{023515} (\bibinfo{year}{2013}), \eprint{1212.0868}.

\bibitem[{\citenamefont{{Senatore}}(2015)}]{SENAT15}
\bibinfo{author}{\bibfnamefont{L.}~\bibnamefont{{Senatore}}},
  \bibinfo{journal}{\jcap} \textbf{\bibinfo{volume}{2015}}, \bibinfo{eid}{007}
  (\bibinfo{year}{2015}), \eprint{1406.7843}.

\bibitem[{\citenamefont{{Mirbabayi} et~al.}(2015)\citenamefont{{Mirbabayi},
  {Schmidt}, and {Zaldarriaga}}}]{MISCZA15}
\bibinfo{author}{\bibfnamefont{M.}~\bibnamefont{{Mirbabayi}}},
  \bibinfo{author}{\bibfnamefont{F.}~\bibnamefont{{Schmidt}}},
  \bibnamefont{and}
  \bibinfo{author}{\bibfnamefont{M.}~\bibnamefont{{Zaldarriaga}}},
  \bibinfo{journal}{\jcap} \textbf{\bibinfo{volume}{2015}},
  \bibinfo{pages}{030} (\bibinfo{year}{2015}), \eprint{1412.5169}.

\bibitem[{\citenamefont{{Angulo} et~al.}(2015)\citenamefont{{Angulo},
  {Fasiello}, {Senatore}, and {Vlah}}}]{ANFAET15}
\bibinfo{author}{\bibfnamefont{R.}~\bibnamefont{{Angulo}}},
  \bibinfo{author}{\bibfnamefont{M.}~\bibnamefont{{Fasiello}}},
  \bibinfo{author}{\bibfnamefont{L.}~\bibnamefont{{Senatore}}},
  \bibnamefont{and} \bibinfo{author}{\bibfnamefont{Z.}~\bibnamefont{{Vlah}}},
  \bibinfo{journal}{\jcap} \textbf{\bibinfo{volume}{2015}},
  \bibinfo{pages}{029} (\bibinfo{year}{2015}), \eprint{1503.08826}.

\bibitem[{\citenamefont{{Ivanov} et~al.}(2020)\citenamefont{{Ivanov},
  {Simonovi{\'c}}, and {Zaldarriaga}}}]{IVSIZA20}
\bibinfo{author}{\bibfnamefont{M.~M.} \bibnamefont{{Ivanov}}},
  \bibinfo{author}{\bibfnamefont{M.}~\bibnamefont{{Simonovi{\'c}}}},
  \bibnamefont{and}
  \bibinfo{author}{\bibfnamefont{M.}~\bibnamefont{{Zaldarriaga}}},
  \bibinfo{journal}{\jcap} \textbf{\bibinfo{volume}{2020}}, \bibinfo{eid}{042}
  (\bibinfo{year}{2020}), \eprint{1909.05277}.

\bibitem[{\citenamefont{{d'Amico} et~al.}(2020)\citenamefont{{d'Amico},
  {Gleyzes}, {Kokron}, {Markovic}, {Senatore}, {Zhang}, {Beutler}, and
  {Gil-Mar{\'\i}n}}}]{DAGLET20}
\bibinfo{author}{\bibfnamefont{G.}~\bibnamefont{{d'Amico}}},
  \bibinfo{author}{\bibfnamefont{J.}~\bibnamefont{{Gleyzes}}},
  \bibinfo{author}{\bibfnamefont{N.}~\bibnamefont{{Kokron}}},
  \bibinfo{author}{\bibfnamefont{K.}~\bibnamefont{{Markovic}}},
  \bibinfo{author}{\bibfnamefont{L.}~\bibnamefont{{Senatore}}},
  \bibinfo{author}{\bibfnamefont{P.}~\bibnamefont{{Zhang}}},
  \bibinfo{author}{\bibfnamefont{F.}~\bibnamefont{{Beutler}}},
  \bibnamefont{and}
  \bibinfo{author}{\bibfnamefont{H.}~\bibnamefont{{Gil-Mar{\'\i}n}}},
  \bibinfo{journal}{\jcap} \textbf{\bibinfo{volume}{2020}}, \bibinfo{eid}{005}
  (\bibinfo{year}{2020}), \eprint{1909.05271}.

\bibitem[{\citenamefont{{Eisenstein} et~al.}(2001)\citenamefont{{Eisenstein},
  {Annis}, {Gunn}, {Szalay}, {Connolly}, {Nichol}, {Bahcall}, {Bernardi},
  {Burles}, {Castander} et~al.}}]{EIANET01}
\bibinfo{author}{\bibfnamefont{D.~J.} \bibnamefont{{Eisenstein}}},
  \bibinfo{author}{\bibfnamefont{J.}~\bibnamefont{{Annis}}},
  \bibinfo{author}{\bibfnamefont{J.~E.} \bibnamefont{{Gunn}}},
  \bibinfo{author}{\bibfnamefont{A.~S.} \bibnamefont{{Szalay}}},
  \bibinfo{author}{\bibfnamefont{A.~J.} \bibnamefont{{Connolly}}},
  \bibinfo{author}{\bibfnamefont{R.~C.} \bibnamefont{{Nichol}}},
  \bibinfo{author}{\bibfnamefont{N.~A.} \bibnamefont{{Bahcall}}},
  \bibinfo{author}{\bibfnamefont{M.}~\bibnamefont{{Bernardi}}},
  \bibinfo{author}{\bibfnamefont{S.}~\bibnamefont{{Burles}}},
  \bibinfo{author}{\bibfnamefont{F.~J.} \bibnamefont{{Castander}}},
  \bibnamefont{et~al.}, \bibinfo{journal}{\aj} \textbf{\bibinfo{volume}{122}},
  \bibinfo{pages}{2267} (\bibinfo{year}{2001}),
  \eprint{arXiv:astro-ph/0108153}.

\bibitem[{\citenamefont{{White} et~al.}(2011)\citenamefont{{White}, {Blanton},
  {Bolton}, {Schlegel}, {Tinker}, {Berlind}, {da Costa}, {Kazin}, {Lin}, {Maia}
  et~al.}}]{WHBLET11}
\bibinfo{author}{\bibfnamefont{M.}~\bibnamefont{{White}}},
  \bibinfo{author}{\bibfnamefont{M.}~\bibnamefont{{Blanton}}},
  \bibinfo{author}{\bibfnamefont{A.}~\bibnamefont{{Bolton}}},
  \bibinfo{author}{\bibfnamefont{D.}~\bibnamefont{{Schlegel}}},
  \bibinfo{author}{\bibfnamefont{J.}~\bibnamefont{{Tinker}}},
  \bibinfo{author}{\bibfnamefont{A.}~\bibnamefont{{Berlind}}},
  \bibinfo{author}{\bibfnamefont{L.}~\bibnamefont{{da Costa}}},
  \bibinfo{author}{\bibfnamefont{E.}~\bibnamefont{{Kazin}}},
  \bibinfo{author}{\bibfnamefont{Y.-T.} \bibnamefont{{Lin}}},
  \bibinfo{author}{\bibfnamefont{M.}~\bibnamefont{{Maia}}},
  \bibnamefont{et~al.}, \bibinfo{journal}{\apj} \textbf{\bibinfo{volume}{728}},
  \bibinfo{pages}{126} (\bibinfo{year}{2011}), \eprint{1010.4915}.

\bibitem[{\citenamefont{{Yoo}}(2010)}]{YOO10}
\bibinfo{author}{\bibfnamefont{J.}~\bibnamefont{{Yoo}}},
  \bibinfo{journal}{\prd} \textbf{\bibinfo{volume}{82}},
  \bibinfo{pages}{083508} (\bibinfo{year}{2010}), \eprint{arXiv:1009.3021}.

\bibitem[{\citenamefont{{Yoo} et~al.}(2012)\citenamefont{{Yoo}, {Hamaus},
  {Seljak}, and {Zaldarriaga}}}]{YOHAET12}
\bibinfo{author}{\bibfnamefont{J.}~\bibnamefont{{Yoo}}},
  \bibinfo{author}{\bibfnamefont{N.}~\bibnamefont{{Hamaus}}},
  \bibinfo{author}{\bibfnamefont{U.}~\bibnamefont{{Seljak}}}, \bibnamefont{and}
  \bibinfo{author}{\bibfnamefont{M.}~\bibnamefont{{Zaldarriaga}}},
  \bibinfo{journal}{\prd} \textbf{\bibinfo{volume}{86}}, \bibinfo{eid}{063514}
  (\bibinfo{year}{2012}), \eprint{1206.5809}.

\bibitem[{\citenamefont{{Dalal} et~al.}(2008)\citenamefont{{Dalal}, {Dor{\'e}},
  {Huterer}, and {Shirokov}}}]{DADOET08}
\bibinfo{author}{\bibfnamefont{N.}~\bibnamefont{{Dalal}}},
  \bibinfo{author}{\bibfnamefont{O.}~\bibnamefont{{Dor{\'e}}}},
  \bibinfo{author}{\bibfnamefont{D.}~\bibnamefont{{Huterer}}},
  \bibnamefont{and}
  \bibinfo{author}{\bibfnamefont{A.}~\bibnamefont{{Shirokov}}},
  \bibinfo{journal}{\prd} \textbf{\bibinfo{volume}{77}},
  \bibinfo{pages}{123514} (\bibinfo{year}{2008}), \eprint{0710.4560}.

\bibitem[{\citenamefont{{Slosar} et~al.}(2008)\citenamefont{{Slosar}, {Hirata},
  {Seljak}, {Ho}, and {Padmanabhan}}}]{SLHIET08}
\bibinfo{author}{\bibfnamefont{A.}~\bibnamefont{{Slosar}}},
  \bibinfo{author}{\bibfnamefont{C.}~\bibnamefont{{Hirata}}},
  \bibinfo{author}{\bibfnamefont{U.}~\bibnamefont{{Seljak}}},
  \bibinfo{author}{\bibfnamefont{S.}~\bibnamefont{{Ho}}}, \bibnamefont{and}
  \bibinfo{author}{\bibfnamefont{N.}~\bibnamefont{{Padmanabhan}}},
  \bibinfo{journal}{\jcap} \textbf{\bibinfo{volume}{8}}, \bibinfo{pages}{31}
  (\bibinfo{year}{2008}), \eprint{0805.3580}.

\bibitem[{\citenamefont{{Matarrese} and {Verde}}(2008)}]{MAVE08}
\bibinfo{author}{\bibfnamefont{S.}~\bibnamefont{{Matarrese}}} \bibnamefont{and}
  \bibinfo{author}{\bibfnamefont{L.}~\bibnamefont{{Verde}}},
  \bibinfo{journal}{\apjl} \textbf{\bibinfo{volume}{677}}, \bibinfo{pages}{L77}
  (\bibinfo{year}{2008}), \eprint{0801.4826}.

\bibitem[{\citenamefont{{Stubbs} et~al.}(2004)\citenamefont{{Stubbs},
  {Sweeney}, {Tyson}, and {LSST Collaboration}}}]{LSST04}
\bibinfo{author}{\bibfnamefont{C.~W.} \bibnamefont{{Stubbs}}},
  \bibinfo{author}{\bibfnamefont{D.}~\bibnamefont{{Sweeney}}},
  \bibinfo{author}{\bibfnamefont{J.~A.} \bibnamefont{{Tyson}}},
  \bibnamefont{and} \bibinfo{author}{\bibnamefont{{LSST Collaboration}}}, in
  \emph{\bibinfo{booktitle}{American Astronomical Society Meeting Abstracts}}
  (\bibinfo{year}{2004}), vol.~\bibinfo{volume}{36} of
  \emph{\bibinfo{series}{Bulletin of the American Astronomical Society}}, p.
  \bibinfo{pages}{108.02}.

\bibitem[{\citenamefont{{Laureijs} et~al.}(2011)\citenamefont{{Laureijs},
  {Amiaux}, {Arduini}, {Augu{\`e}res}, {Brinchmann}, {Cole}, {Cropper}, {Dabin}
  et~al.}}]{EUCLID11}
\bibinfo{author}{\bibfnamefont{R.}~\bibnamefont{{Laureijs}}},
  \bibinfo{author}{\bibfnamefont{J.}~\bibnamefont{{Amiaux}}},
  \bibinfo{author}{\bibfnamefont{S.}~\bibnamefont{{Arduini}}},
  \bibinfo{author}{\bibfnamefont{J.~.} \bibnamefont{{Augu{\`e}res}}},
  \bibinfo{author}{\bibfnamefont{J.}~\bibnamefont{{Brinchmann}}},
  \bibinfo{author}{\bibfnamefont{R.}~\bibnamefont{{Cole}}},
  \bibinfo{author}{\bibfnamefont{M.}~\bibnamefont{{Cropper}}},
  \bibinfo{author}{\bibfnamefont{C.}~\bibnamefont{{Dabin}}},
  \bibnamefont{et~al.}, \bibinfo{journal}{ArXiv e-prints}
  (\bibinfo{year}{2011}), \eprint{1110.3193}.

\bibitem[{\citenamefont{{Green} et~al.}(2012)\citenamefont{{Green},
  {Schechter}, {Baltay}, {Bean}, {Bennett}, {Brown}, {Conselice}, {Donahue}
  et~al.}}]{WFIRST12}
\bibinfo{author}{\bibfnamefont{J.}~\bibnamefont{{Green}}},
  \bibinfo{author}{\bibfnamefont{P.}~\bibnamefont{{Schechter}}},
  \bibinfo{author}{\bibfnamefont{C.}~\bibnamefont{{Baltay}}},
  \bibinfo{author}{\bibfnamefont{R.}~\bibnamefont{{Bean}}},
  \bibinfo{author}{\bibfnamefont{D.}~\bibnamefont{{Bennett}}},
  \bibinfo{author}{\bibfnamefont{R.}~\bibnamefont{{Brown}}},
  \bibinfo{author}{\bibfnamefont{C.}~\bibnamefont{{Conselice}}},
  \bibinfo{author}{\bibfnamefont{M.}~\bibnamefont{{Donahue}}},
  \bibnamefont{et~al.} (\bibinfo{year}{2012}), \eprint{1208.4012}.

\bibitem[{\citenamefont{{Cosmic Visions 21 cm Collaboration}
  et~al.}(2018)\citenamefont{{Cosmic Visions 21 cm Collaboration}, {Ansari},
  {Arena}, {Bandura}, {Bull}, {Castorina}, {Chang}, {Chen} et~al.}}]{ANARET18}
\bibinfo{author}{\bibnamefont{{Cosmic Visions 21 cm Collaboration}}},
  \bibinfo{author}{\bibfnamefont{R.}~\bibnamefont{{Ansari}}},
  \bibinfo{author}{\bibfnamefont{E.~J.} \bibnamefont{{Arena}}},
  \bibinfo{author}{\bibfnamefont{K.}~\bibnamefont{{Bandura}}},
  \bibinfo{author}{\bibfnamefont{P.}~\bibnamefont{{Bull}}},
  \bibinfo{author}{\bibfnamefont{E.}~\bibnamefont{{Castorina}}},
  \bibinfo{author}{\bibfnamefont{T.-C.} \bibnamefont{{Chang}}},
  \bibinfo{author}{\bibnamefont{{Chen}}}, \bibnamefont{et~al.},
  \bibinfo{journal}{arXiv e-prints} \bibinfo{eid}{arXiv:1810.09572}
  (\bibinfo{year}{2018}), \eprint{1810.09572}.

\bibitem[{\citenamefont{{Mueller} et~al.}(2022)\citenamefont{{Mueller},
  {Rezaie}, {Percival}, {Ross} et~al.}}]{MUREET22}
\bibinfo{author}{\bibfnamefont{E.-M.} \bibnamefont{{Mueller}}},
  \bibinfo{author}{\bibfnamefont{M.}~\bibnamefont{{Rezaie}}},
  \bibinfo{author}{\bibfnamefont{W.~J.} \bibnamefont{{Percival}}},
  \bibinfo{author}{\bibfnamefont{A.~J.} \bibnamefont{{Ross}}},
  \bibnamefont{et~al.}, \bibinfo{journal}{\mnras}
  \textbf{\bibinfo{volume}{514}}, \bibinfo{pages}{3396} (\bibinfo{year}{2022}).

\bibitem[{\citenamefont{{Cabass} et~al.}(2022)\citenamefont{{Cabass}, {Ivanov},
  {Philcox}, {Simonovi{\'c}}, and {Zaldarriaga}}}]{CAIVET22}
\bibinfo{author}{\bibfnamefont{G.}~\bibnamefont{{Cabass}}},
  \bibinfo{author}{\bibfnamefont{M.~M.} \bibnamefont{{Ivanov}}},
  \bibinfo{author}{\bibfnamefont{O.~H.~E.} \bibnamefont{{Philcox}}},
  \bibinfo{author}{\bibfnamefont{M.}~\bibnamefont{{Simonovi{\'c}}}},
  \bibnamefont{and}
  \bibinfo{author}{\bibfnamefont{M.}~\bibnamefont{{Zaldarriaga}}},
  \bibinfo{journal}{\prd} \textbf{\bibinfo{volume}{106}}, \bibinfo{eid}{043506}
  (\bibinfo{year}{2022}), \eprint{2204.01781}.

\bibitem[{\citenamefont{{D'Amico} et~al.}(2022)\citenamefont{{D'Amico},
  {Lewandowski}, {Senatore}, and {Zhang}}}]{DALEET22}
\bibinfo{author}{\bibfnamefont{G.}~\bibnamefont{{D'Amico}}},
  \bibinfo{author}{\bibfnamefont{M.}~\bibnamefont{{Lewandowski}}},
  \bibinfo{author}{\bibfnamefont{L.}~\bibnamefont{{Senatore}}},
  \bibnamefont{and} \bibinfo{author}{\bibfnamefont{P.}~\bibnamefont{{Zhang}}},
  \bibinfo{journal}{arXiv e-prints} \bibinfo{eid}{arXiv:2201.11518}
  (\bibinfo{year}{2022}), \eprint{2201.11518}.

\bibitem[{\citenamefont{{Bruni}
  et~al.}(2014{\natexlab{a}})\citenamefont{{Bruni}, {Hidalgo}, {Meures}, and
  {Wands}}}]{BRHIET14}
\bibinfo{author}{\bibfnamefont{M.}~\bibnamefont{{Bruni}}},
  \bibinfo{author}{\bibfnamefont{J.~C.} \bibnamefont{{Hidalgo}}},
  \bibinfo{author}{\bibfnamefont{N.}~\bibnamefont{{Meures}}}, \bibnamefont{and}
  \bibinfo{author}{\bibfnamefont{D.}~\bibnamefont{{Wands}}},
  \bibinfo{journal}{\apj} \textbf{\bibinfo{volume}{785}}, \bibinfo{pages}{2}
  (\bibinfo{year}{2014}{\natexlab{a}}), \eprint{1307.1478}.

\bibitem[{\citenamefont{{Bruni}
  et~al.}(2014{\natexlab{b}})\citenamefont{{Bruni}, {Hidalgo}, and
  {Wands}}}]{BRHIWA14}
\bibinfo{author}{\bibfnamefont{M.}~\bibnamefont{{Bruni}}},
  \bibinfo{author}{\bibfnamefont{J.~C.} \bibnamefont{{Hidalgo}}},
  \bibnamefont{and} \bibinfo{author}{\bibfnamefont{D.}~\bibnamefont{{Wands}}},
  \bibinfo{journal}{\apjl} \textbf{\bibinfo{volume}{794}}, \bibinfo{eid}{L11}
  (\bibinfo{year}{2014}{\natexlab{b}}), \eprint{1405.7006}.

\bibitem[{\citenamefont{{Villa} et~al.}(2014)\citenamefont{{Villa}, {Verde},
  and {Matarrese}}}]{VIVEMA14}
\bibinfo{author}{\bibfnamefont{E.}~\bibnamefont{{Villa}}},
  \bibinfo{author}{\bibfnamefont{L.}~\bibnamefont{{Verde}}}, \bibnamefont{and}
  \bibinfo{author}{\bibfnamefont{S.}~\bibnamefont{{Matarrese}}},
  \bibinfo{journal}{Classical and Quantum Gravity}
  \textbf{\bibinfo{volume}{31}}, \bibinfo{eid}{234005} (\bibinfo{year}{2014}),
  \eprint{1409.4738}.

\bibitem[{\citenamefont{{Camera} et~al.}(2015)\citenamefont{{Camera},
  {Maartens}, and {Santos}}}]{CAMASA15}
\bibinfo{author}{\bibfnamefont{S.}~\bibnamefont{{Camera}}},
  \bibinfo{author}{\bibfnamefont{R.}~\bibnamefont{{Maartens}}},
  \bibnamefont{and} \bibinfo{author}{\bibfnamefont{M.~G.}
  \bibnamefont{{Santos}}}, \bibinfo{journal}{\mnras}
  \textbf{\bibinfo{volume}{451}}, \bibinfo{pages}{L80} (\bibinfo{year}{2015}),
  \eprint{1412.4781}.

\bibitem[{\citenamefont{{Matarrese} et~al.}(2021)\citenamefont{{Matarrese},
  {Pilo}, and {Rollo}}}]{MAPIRO21}
\bibinfo{author}{\bibfnamefont{S.}~\bibnamefont{{Matarrese}}},
  \bibinfo{author}{\bibfnamefont{L.}~\bibnamefont{{Pilo}}}, \bibnamefont{and}
  \bibinfo{author}{\bibfnamefont{R.}~\bibnamefont{{Rollo}}},
  \bibinfo{journal}{\jcap} \textbf{\bibinfo{volume}{2021}}, \bibinfo{eid}{062}
  (\bibinfo{year}{2021}), \eprint{2007.08877}.

\bibitem[{\citenamefont{{Castorina} and {Di Dio}}(2022)}]{CADI22}
\bibinfo{author}{\bibfnamefont{E.}~\bibnamefont{{Castorina}}} \bibnamefont{and}
  \bibinfo{author}{\bibfnamefont{E.}~\bibnamefont{{Di Dio}}},
  \bibinfo{journal}{\jcap} \textbf{\bibinfo{volume}{2022}}, \bibinfo{eid}{061}
  (\bibinfo{year}{2022}), \eprint{2106.08857}.

\bibitem[{\citenamefont{{Foglieni} et~al.}(2023)\citenamefont{{Foglieni},
  {Pantiri}, {Di Dio}, and {Castorina}}}]{FOPAET23}
\bibinfo{author}{\bibfnamefont{M.}~\bibnamefont{{Foglieni}}},
  \bibinfo{author}{\bibfnamefont{M.}~\bibnamefont{{Pantiri}}},
  \bibinfo{author}{\bibfnamefont{E.}~\bibnamefont{{Di Dio}}}, \bibnamefont{and}
  \bibinfo{author}{\bibfnamefont{E.}~\bibnamefont{{Castorina}}},
  \bibinfo{journal}{arXiv e-prints} \bibinfo{eid}{arXiv:2303.03142}
  (\bibinfo{year}{2023}), \eprint{2303.03142}.

\bibitem[{\citenamefont{{Mitsou} and {Yoo}}(2020)}]{MIYO20}
\bibinfo{author}{\bibfnamefont{E.}~\bibnamefont{{Mitsou}}} \bibnamefont{and}
  \bibinfo{author}{\bibfnamefont{J.}~\bibnamefont{{Yoo}}},
  \bibinfo{journal}{Springer Briefs in Physics} \bibinfo{eid}{arXiv:1908.10757}
  (\bibinfo{year}{2020}), \eprint{1908.10757}.

\bibitem[{\citenamefont{{Yoo} et~al.}(2022)\citenamefont{{Yoo}, {Grimm}, and
  {Mitsou}}}]{YOGRMI22}
\bibinfo{author}{\bibfnamefont{J.}~\bibnamefont{{Yoo}}},
  \bibinfo{author}{\bibfnamefont{N.}~\bibnamefont{{Grimm}}}, \bibnamefont{and}
  \bibinfo{author}{\bibfnamefont{E.}~\bibnamefont{{Mitsou}}},
  \bibinfo{journal}{\jcap} \textbf{\bibinfo{volume}{2022}}, \bibinfo{eid}{050}
  (\bibinfo{year}{2022}), \eprint{2204.03002}.

\bibitem[{\citenamefont{{Challinor} and {Lewis}}(2011)}]{CHLE11}
\bibinfo{author}{\bibfnamefont{A.}~\bibnamefont{{Challinor}}} \bibnamefont{and}
  \bibinfo{author}{\bibfnamefont{A.}~\bibnamefont{{Lewis}}},
  \bibinfo{journal}{\prd} \textbf{\bibinfo{volume}{84}}, \bibinfo{eid}{043516}
  (\bibinfo{year}{2011}), \eprint{arXiv:1105.5292}.

\bibitem[{\citenamefont{{Jeong} et~al.}(2012)\citenamefont{{Jeong}, {Schmidt},
  and {Hirata}}}]{JESCHI12}
\bibinfo{author}{\bibfnamefont{D.}~\bibnamefont{{Jeong}}},
  \bibinfo{author}{\bibfnamefont{F.}~\bibnamefont{{Schmidt}}},
  \bibnamefont{and} \bibinfo{author}{\bibfnamefont{C.~M.}
  \bibnamefont{{Hirata}}}, \bibinfo{journal}{\prd}
  \textbf{\bibinfo{volume}{85}}, \bibinfo{pages}{023504}
  (\bibinfo{year}{2012}), \eprint{arXiv:1107.5427}.

\bibitem[{\citenamefont{{Jeong} and {Schmidt}}(2014)}]{JESC13}
\bibinfo{author}{\bibfnamefont{D.}~\bibnamefont{{Jeong}}} \bibnamefont{and}
  \bibinfo{author}{\bibfnamefont{F.}~\bibnamefont{{Schmidt}}},
  \bibinfo{journal}{\prd} \textbf{\bibinfo{volume}{89}},
  \bibinfo{pages}{043519} (\bibinfo{year}{2014}), \eprint{1305.1299}.

\bibitem[{\citenamefont{{Yoo}}(2014{\natexlab{a}})}]{YOO14b}
\bibinfo{author}{\bibfnamefont{J.}~\bibnamefont{{Yoo}}},
  \bibinfo{journal}{\prd} \textbf{\bibinfo{volume}{90}},
  \bibinfo{pages}{123507} (\bibinfo{year}{2014}{\natexlab{a}}),
  \eprint{1408.5137}.

\bibitem[{\citenamefont{{Yoo} et~al.}(2018)\citenamefont{{Yoo}, {Grimm},
  {Mitsou}, {Amara}, and {Refregier}}}]{YOGRET18}
\bibinfo{author}{\bibfnamefont{J.}~\bibnamefont{{Yoo}}},
  \bibinfo{author}{\bibfnamefont{N.}~\bibnamefont{{Grimm}}},
  \bibinfo{author}{\bibfnamefont{E.}~\bibnamefont{{Mitsou}}},
  \bibinfo{author}{\bibfnamefont{A.}~\bibnamefont{{Amara}}}, \bibnamefont{and}
  \bibinfo{author}{\bibfnamefont{A.}~\bibnamefont{{Refregier}}},
  \bibinfo{journal}{\jcap} \textbf{\bibinfo{volume}{4}}, \bibinfo{eid}{029}
  (\bibinfo{year}{2018}), \eprint{1802.03403}.

\bibitem[{\citenamefont{{Planck Collaboration}
  et~al.}(2020)\citenamefont{{Planck Collaboration}, {Aghanim}, {Akrami},
  {Ashdown}, {Aumont}, {Baccigalupi}, {Ballardini}, {Banday}, {Barreiro},
  {Bartolo} et~al.}}]{PLANCKcos18}
\bibinfo{author}{\bibnamefont{{Planck Collaboration}}},
  \bibinfo{author}{\bibfnamefont{N.}~\bibnamefont{{Aghanim}}},
  \bibinfo{author}{\bibfnamefont{Y.}~\bibnamefont{{Akrami}}},
  \bibinfo{author}{\bibfnamefont{M.}~\bibnamefont{{Ashdown}}},
  \bibinfo{author}{\bibfnamefont{J.}~\bibnamefont{{Aumont}}},
  \bibinfo{author}{\bibfnamefont{C.}~\bibnamefont{{Baccigalupi}}},
  \bibinfo{author}{\bibfnamefont{M.}~\bibnamefont{{Ballardini}}},
  \bibinfo{author}{\bibfnamefont{A.~J.} \bibnamefont{{Banday}}},
  \bibinfo{author}{\bibfnamefont{R.~B.} \bibnamefont{{Barreiro}}},
  \bibinfo{author}{\bibfnamefont{N.}~\bibnamefont{{Bartolo}}},
  \bibnamefont{et~al.}, \bibinfo{journal}{\aap} \textbf{\bibinfo{volume}{641}},
  \bibinfo{eid}{A6} (\bibinfo{year}{2020}), \eprint{1807.06209}.

\bibitem[{\citenamefont{{Fidler} et~al.}(2015)\citenamefont{{Fidler}, {Rampf},
  {Tram}, {Crittenden}, {Koyama}, and {Wands}}}]{FIRAET15}
\bibinfo{author}{\bibfnamefont{C.}~\bibnamefont{{Fidler}}},
  \bibinfo{author}{\bibfnamefont{C.}~\bibnamefont{{Rampf}}},
  \bibinfo{author}{\bibfnamefont{T.}~\bibnamefont{{Tram}}},
  \bibinfo{author}{\bibfnamefont{R.}~\bibnamefont{{Crittenden}}},
  \bibinfo{author}{\bibfnamefont{K.}~\bibnamefont{{Koyama}}}, \bibnamefont{and}
  \bibinfo{author}{\bibfnamefont{D.}~\bibnamefont{{Wands}}},
  \bibinfo{journal}{\prd} \textbf{\bibinfo{volume}{92}},
  \bibinfo{pages}{123517} (\bibinfo{year}{2015}), \eprint{1505.04756}.

\bibitem[{\citenamefont{{Fidler} et~al.}(2016)\citenamefont{{Fidler}, {Tram},
  {Rampf}, {Crittenden}, {Koyama}, and {Wands}}}]{FITRET16}
\bibinfo{author}{\bibfnamefont{C.}~\bibnamefont{{Fidler}}},
  \bibinfo{author}{\bibfnamefont{T.}~\bibnamefont{{Tram}}},
  \bibinfo{author}{\bibfnamefont{C.}~\bibnamefont{{Rampf}}},
  \bibinfo{author}{\bibfnamefont{R.}~\bibnamefont{{Crittenden}}},
  \bibinfo{author}{\bibfnamefont{K.}~\bibnamefont{{Koyama}}}, \bibnamefont{and}
  \bibinfo{author}{\bibfnamefont{D.}~\bibnamefont{{Wands}}},
  \bibinfo{journal}{Journal of Cosmology and Astro-Particle Physics}
  \textbf{\bibinfo{volume}{2016}}, \bibinfo{pages}{031} (\bibinfo{year}{2016}),
  \eprint{1606.05588}.

\bibitem[{\citenamefont{{Fidler} et~al.}(2017)\citenamefont{{Fidler}, {Tram},
  {Rampf}, {Crittenden}, {Koyama}, and {Wands}}}]{FITRET17}
\bibinfo{author}{\bibfnamefont{C.}~\bibnamefont{{Fidler}}},
  \bibinfo{author}{\bibfnamefont{T.}~\bibnamefont{{Tram}}},
  \bibinfo{author}{\bibfnamefont{C.}~\bibnamefont{{Rampf}}},
  \bibinfo{author}{\bibfnamefont{R.}~\bibnamefont{{Crittenden}}},
  \bibinfo{author}{\bibfnamefont{K.}~\bibnamefont{{Koyama}}}, \bibnamefont{and}
  \bibinfo{author}{\bibfnamefont{D.}~\bibnamefont{{Wands}}},
  \bibinfo{journal}{Journal of Cosmology and Astro-Particle Physics}
  \textbf{\bibinfo{volume}{2017}}, \bibinfo{eid}{022} (\bibinfo{year}{2017}),
  \eprint{1708.07769}.

\bibitem[{\citenamefont{{Adamek} and {Fidler}}(2019)}]{ADFI19}
\bibinfo{author}{\bibfnamefont{J.}~\bibnamefont{{Adamek}}} \bibnamefont{and}
  \bibinfo{author}{\bibfnamefont{C.}~\bibnamefont{{Fidler}}},
  \bibinfo{journal}{arXiv e-prints} p. \bibinfo{pages}{arXiv:1905.11721}
  (\bibinfo{year}{2019}), \eprint{1905.11721}.

\bibitem[{\citenamefont{{Rampf}}(2014)}]{RAMPF14}
\bibinfo{author}{\bibfnamefont{C.}~\bibnamefont{{Rampf}}},
  \bibinfo{journal}{\prd} \textbf{\bibinfo{volume}{89}}, \bibinfo{eid}{063509}
  (\bibinfo{year}{2014}), \eprint{1307.1725}.

\bibitem[{\citenamefont{{Rampf} and {Wiegand}}(2014)}]{RAWI14}
\bibinfo{author}{\bibfnamefont{C.}~\bibnamefont{{Rampf}}} \bibnamefont{and}
  \bibinfo{author}{\bibfnamefont{A.}~\bibnamefont{{Wiegand}}},
  \bibinfo{journal}{\prd} \textbf{\bibinfo{volume}{90}}, \bibinfo{eid}{123503}
  (\bibinfo{year}{2014}), \eprint{1409.2688}.

\bibitem[{\citenamefont{{Koyama} et~al.}(2018)\citenamefont{{Koyama}, {Umeh},
  {Maartens}, and {Bertacca}}}]{KOUMET18}
\bibinfo{author}{\bibfnamefont{K.}~\bibnamefont{{Koyama}}},
  \bibinfo{author}{\bibfnamefont{O.}~\bibnamefont{{Umeh}}},
  \bibinfo{author}{\bibfnamefont{R.}~\bibnamefont{{Maartens}}},
  \bibnamefont{and}
  \bibinfo{author}{\bibfnamefont{D.}~\bibnamefont{{Bertacca}}},
  \bibinfo{journal}{\jcap} \textbf{\bibinfo{volume}{2018}}, \bibinfo{eid}{050}
  (\bibinfo{year}{2018}), \eprint{1805.09189}.

\bibitem[{\citenamefont{{Umeh} et~al.}(2019)\citenamefont{{Umeh}, {Koyama},
  {Maartens}, {Schmidt}, and {Clarkson}}}]{UMKOET19}
\bibinfo{author}{\bibfnamefont{O.}~\bibnamefont{{Umeh}}},
  \bibinfo{author}{\bibfnamefont{K.}~\bibnamefont{{Koyama}}},
  \bibinfo{author}{\bibfnamefont{R.}~\bibnamefont{{Maartens}}},
  \bibinfo{author}{\bibfnamefont{F.}~\bibnamefont{{Schmidt}}},
  \bibnamefont{and}
  \bibinfo{author}{\bibfnamefont{C.}~\bibnamefont{{Clarkson}}},
  \bibinfo{journal}{\jcap} \textbf{\bibinfo{volume}{2019}}, \bibinfo{eid}{020}
  (\bibinfo{year}{2019}), \eprint{1901.07460}.

\bibitem[{\citenamefont{{Umeh} and {Koyama}}(2019)}]{UMKO19}
\bibinfo{author}{\bibfnamefont{O.}~\bibnamefont{{Umeh}}} \bibnamefont{and}
  \bibinfo{author}{\bibfnamefont{K.}~\bibnamefont{{Koyama}}},
  \bibinfo{journal}{\jcap} \textbf{\bibinfo{volume}{2019}}, \bibinfo{eid}{048}
  (\bibinfo{year}{2019}), \eprint{1907.08094}.

\bibitem[{\citenamefont{{Stewart} and {Walker}}(1974)}]{STWA74}
\bibinfo{author}{\bibfnamefont{J.~M.} \bibnamefont{{Stewart}}}
  \bibnamefont{and} \bibinfo{author}{\bibfnamefont{M.}~\bibnamefont{{Walker}}},
  \bibinfo{journal}{Proceedings of the Royal Society of London Series A}
  \textbf{\bibinfo{volume}{341}}, \bibinfo{pages}{49} (\bibinfo{year}{1974}).

\bibitem[{\citenamefont{{Bruni} et~al.}(1997)\citenamefont{{Bruni},
  {Matarrese}, {Mollerach}, and {Sonego}}}]{BRMAET97}
\bibinfo{author}{\bibfnamefont{M.}~\bibnamefont{{Bruni}}},
  \bibinfo{author}{\bibfnamefont{S.}~\bibnamefont{{Matarrese}}},
  \bibinfo{author}{\bibfnamefont{S.}~\bibnamefont{{Mollerach}}},
  \bibnamefont{and} \bibinfo{author}{\bibfnamefont{S.}~\bibnamefont{{Sonego}}},
  \bibinfo{journal}{Classical and Quantum Gravity}
  \textbf{\bibinfo{volume}{14}}, \bibinfo{pages}{2585} (\bibinfo{year}{1997}),
  \eprint{gr-qc/9609040}.

\bibitem[{\citenamefont{{Yoo} and {Durrer}}(2017)}]{YODU17}
\bibinfo{author}{\bibfnamefont{J.}~\bibnamefont{{Yoo}}} \bibnamefont{and}
  \bibinfo{author}{\bibfnamefont{R.}~\bibnamefont{{Durrer}}},
  \bibinfo{journal}{\jcap} \textbf{\bibinfo{volume}{9}}, \bibinfo{eid}{016}
  (\bibinfo{year}{2017}), \eprint{1705.05839}.

\bibitem[{\citenamefont{{Maldacena}}(2003)}]{MALDA03}
\bibinfo{author}{\bibfnamefont{J.}~\bibnamefont{{Maldacena}}},
  \bibinfo{journal}{\jhep} \textbf{\bibinfo{volume}{5}}, \bibinfo{pages}{13}
  (\bibinfo{year}{2003}), \eprint{astro-ph/0210603}.

\bibitem[{\citenamefont{{Creminelli} and {Zaldarriaga}}(2004)}]{CRZA04}
\bibinfo{author}{\bibfnamefont{P.}~\bibnamefont{{Creminelli}}}
  \bibnamefont{and}
  \bibinfo{author}{\bibfnamefont{M.}~\bibnamefont{{Zaldarriaga}}},
  \bibinfo{journal}{\jcap} \textbf{\bibinfo{volume}{10}}, \bibinfo{pages}{6}
  (\bibinfo{year}{2004}), \eprint{arXiv:astro-ph/0407059}.

\bibitem[{\citenamefont{{Chen} et~al.}(2007)\citenamefont{{Chen}, {Huang},
  {Kachru}, and {Shiu}}}]{CHHUET07}
\bibinfo{author}{\bibfnamefont{X.}~\bibnamefont{{Chen}}},
  \bibinfo{author}{\bibfnamefont{M.-x.} \bibnamefont{{Huang}}},
  \bibinfo{author}{\bibfnamefont{S.}~\bibnamefont{{Kachru}}}, \bibnamefont{and}
  \bibinfo{author}{\bibfnamefont{G.}~\bibnamefont{{Shiu}}},
  \bibinfo{journal}{\jcap} \textbf{\bibinfo{volume}{2007}}, \bibinfo{eid}{002}
  (\bibinfo{year}{2007}), \eprint{hep-th/0605045}.

\bibitem[{\citenamefont{{Senatore} and {Zaldarriaga}}(2012)}]{SEZA12}
\bibinfo{author}{\bibfnamefont{L.}~\bibnamefont{{Senatore}}} \bibnamefont{and}
  \bibinfo{author}{\bibfnamefont{M.}~\bibnamefont{{Zaldarriaga}}},
  \bibinfo{journal}{\jcap} \textbf{\bibinfo{volume}{2012}}, \bibinfo{eid}{001}
  (\bibinfo{year}{2012}), \eprint{1203.6884}.

\bibitem[{\citenamefont{{Mitsou} and {Yoo}}(2022)}]{MIYO22a}
\bibinfo{author}{\bibfnamefont{E.}~\bibnamefont{{Mitsou}}} \bibnamefont{and}
  \bibinfo{author}{\bibfnamefont{J.}~\bibnamefont{{Yoo}}},
  \bibinfo{journal}{Physics Letters B} \textbf{\bibinfo{volume}{828}},
  \bibinfo{eid}{137018} (\bibinfo{year}{2022}), \eprint{2109.13154}.

\bibitem[{\citenamefont{{Szalay}}(1988)}]{SZALA88}
\bibinfo{author}{\bibfnamefont{A.~S.} \bibnamefont{{Szalay}}},
  \bibinfo{journal}{\apj} \textbf{\bibinfo{volume}{333}}, \bibinfo{pages}{21}
  (\bibinfo{year}{1988}).

\bibitem[{\citenamefont{{Fry} and {Gaztanaga}}(1993)}]{FRGA93}
\bibinfo{author}{\bibfnamefont{J.~N.} \bibnamefont{{Fry}}} \bibnamefont{and}
  \bibinfo{author}{\bibfnamefont{E.}~\bibnamefont{{Gaztanaga}}},
  \bibinfo{journal}{\apj} \textbf{\bibinfo{volume}{413}}, \bibinfo{pages}{447}
  (\bibinfo{year}{1993}), \eprint{astro-ph/9302009}.

\bibitem[{\citenamefont{{Fry}}(1996)}]{FRY96}
\bibinfo{author}{\bibfnamefont{J.~N.} \bibnamefont{{Fry}}},
  \bibinfo{journal}{\apjl} \textbf{\bibinfo{volume}{461}},
  \bibinfo{pages}{L65+} (\bibinfo{year}{1996}).

\bibitem[{\citenamefont{{Yoo} et~al.}(2009)\citenamefont{{Yoo}, {Fitzpatrick},
  and {Zaldarriaga}}}]{YOFIZA09}
\bibinfo{author}{\bibfnamefont{J.}~\bibnamefont{{Yoo}}},
  \bibinfo{author}{\bibfnamefont{A.~L.} \bibnamefont{{Fitzpatrick}}},
  \bibnamefont{and}
  \bibinfo{author}{\bibfnamefont{M.}~\bibnamefont{{Zaldarriaga}}},
  \bibinfo{journal}{\prd} \textbf{\bibinfo{volume}{80}},
  \bibinfo{pages}{083514} (\bibinfo{year}{2009}), \eprint{arXiv:0907.0707}.

\bibitem[{\citenamefont{{Bonvin} and {Durrer}}(2011)}]{BODU11}
\bibinfo{author}{\bibfnamefont{C.}~\bibnamefont{{Bonvin}}} \bibnamefont{and}
  \bibinfo{author}{\bibfnamefont{R.}~\bibnamefont{{Durrer}}},
  \bibinfo{journal}{\prd} \textbf{\bibinfo{volume}{84}}, \bibinfo{eid}{063505}
  (\bibinfo{year}{2011}), \eprint{arXiv:1105.5280}.

\bibitem[{\citenamefont{{Yoo}}(2014{\natexlab{b}})}]{YOO14a}
\bibinfo{author}{\bibfnamefont{J.}~\bibnamefont{{Yoo}}},
  \bibinfo{journal}{\cqg} \textbf{\bibinfo{volume}{31}}, \bibinfo{eid}{234001}
  (\bibinfo{year}{2014}{\natexlab{b}}), \eprint{arXiv:1409.3223}.

\bibitem[{\citenamefont{{Biern} and {Yoo}}(2017)}]{BIYO17}
\bibinfo{author}{\bibfnamefont{S.~G.} \bibnamefont{{Biern}}} \bibnamefont{and}
  \bibinfo{author}{\bibfnamefont{J.}~\bibnamefont{{Yoo}}},
  \bibinfo{journal}{\jcap} \bibinfo{eid}{026} (\bibinfo{year}{2017}),
  \eprint{1704.07380}.

\bibitem[{\citenamefont{{Scaccabarozzi}
  et~al.}(2018)\citenamefont{{Scaccabarozzi}, {Yoo}, and {Biern}}}]{SCYOBI18}
\bibinfo{author}{\bibfnamefont{F.}~\bibnamefont{{Scaccabarozzi}}},
  \bibinfo{author}{\bibfnamefont{J.}~\bibnamefont{{Yoo}}}, \bibnamefont{and}
  \bibinfo{author}{\bibfnamefont{S.~G.} \bibnamefont{{Biern}}},
  \bibinfo{journal}{\jcap} \textbf{\bibinfo{volume}{10}}, \bibinfo{pages}{024}
  (\bibinfo{year}{2018}), \eprint{1807.09796}.

\bibitem[{\citenamefont{{Grimm} et~al.}(2020)\citenamefont{{Grimm},
  {Scaccabarozzi}, {Yoo}, {Biern}, and {Gong}}}]{GRSCET20}
\bibinfo{author}{\bibfnamefont{N.}~\bibnamefont{{Grimm}}},
  \bibinfo{author}{\bibfnamefont{F.}~\bibnamefont{{Scaccabarozzi}}},
  \bibinfo{author}{\bibfnamefont{J.}~\bibnamefont{{Yoo}}},
  \bibinfo{author}{\bibfnamefont{S.~G.} \bibnamefont{{Biern}}},
  \bibnamefont{and} \bibinfo{author}{\bibfnamefont{J.-O.}
  \bibnamefont{{Gong}}}, \bibinfo{journal}{\jcap}
  \textbf{\bibinfo{volume}{2020}}, \bibinfo{eid}{064} (\bibinfo{year}{2020}),
  \eprint{2005.06484}.

\bibitem[{\citenamefont{{Baumgartner} and {Yoo}}(2021)}]{BAYO21}
\bibinfo{author}{\bibfnamefont{S.}~\bibnamefont{{Baumgartner}}}
  \bibnamefont{and} \bibinfo{author}{\bibfnamefont{J.}~\bibnamefont{{Yoo}}},
  \bibinfo{journal}{\prd} \textbf{\bibinfo{volume}{103}}, \bibinfo{eid}{063516}
  (\bibinfo{year}{2021}), \eprint{2012.03968}.

\bibitem[{\citenamefont{{Komatsu} and {Spergel}}(2001)}]{KOSP01}
\bibinfo{author}{\bibfnamefont{E.}~\bibnamefont{{Komatsu}}} \bibnamefont{and}
  \bibinfo{author}{\bibfnamefont{D.~N.} \bibnamefont{{Spergel}}},
  \bibinfo{journal}{\prd} \textbf{\bibinfo{volume}{63}}, \bibinfo{eid}{063002}
  (\bibinfo{year}{2001}), \eprint{astro-ph/0005036}.

\bibitem[{\citenamefont{{Bartolo} et~al.}(2004)\citenamefont{{Bartolo},
  {Komatsu}, {Matarrese}, and {Riotto}}}]{BAKOET04}
\bibinfo{author}{\bibfnamefont{N.}~\bibnamefont{{Bartolo}}},
  \bibinfo{author}{\bibfnamefont{E.}~\bibnamefont{{Komatsu}}},
  \bibinfo{author}{\bibfnamefont{S.}~\bibnamefont{{Matarrese}}},
  \bibnamefont{and} \bibinfo{author}{\bibfnamefont{A.}~\bibnamefont{{Riotto}}},
  \bibinfo{journal}{\physrep} \textbf{\bibinfo{volume}{402}},
  \bibinfo{pages}{103} (\bibinfo{year}{2004}), \eprint{arXiv:0406398}.

\bibitem[{\citenamefont{{Seery} and {Lidsey}}(2005)}]{SELI05a}
\bibinfo{author}{\bibfnamefont{D.}~\bibnamefont{{Seery}}} \bibnamefont{and}
  \bibinfo{author}{\bibfnamefont{J.~E.} \bibnamefont{{Lidsey}}},
  \bibinfo{journal}{\jcap} \textbf{\bibinfo{volume}{6}}, \bibinfo{eid}{003}
  (\bibinfo{year}{2005}), \eprint{astro-ph/0503692}.

\bibitem[{\citenamefont{{Byrnes} et~al.}(2006)\citenamefont{{Byrnes}, {Sasaki},
  and {Wands}}}]{BYSAWA06}
\bibinfo{author}{\bibfnamefont{C.~T.} \bibnamefont{{Byrnes}}},
  \bibinfo{author}{\bibfnamefont{M.}~\bibnamefont{{Sasaki}}}, \bibnamefont{and}
  \bibinfo{author}{\bibfnamefont{D.}~\bibnamefont{{Wands}}},
  \bibinfo{journal}{\prd} \textbf{\bibinfo{volume}{74}},
  \bibinfo{pages}{123519} (\bibinfo{year}{2006}),
  \eprint{arXiv:astro-ph/0611075}.

\bibitem[{\citenamefont{{De Felice} and {Tsujikawa}}(2011)}]{DETS11}
\bibinfo{author}{\bibfnamefont{A.}~\bibnamefont{{De Felice}}} \bibnamefont{and}
  \bibinfo{author}{\bibfnamefont{S.}~\bibnamefont{{Tsujikawa}}},
  \bibinfo{journal}{\jcap} \textbf{\bibinfo{volume}{2011}}, \bibinfo{eid}{029}
  (\bibinfo{year}{2011}), \eprint{1103.1172}.

\bibitem[{\citenamefont{{Matarrese} et~al.}(1986)\citenamefont{{Matarrese},
  {Lucchin}, and {Bonometto}}}]{MALUBO86}
\bibinfo{author}{\bibfnamefont{S.}~\bibnamefont{{Matarrese}}},
  \bibinfo{author}{\bibfnamefont{F.}~\bibnamefont{{Lucchin}}},
  \bibnamefont{and} \bibinfo{author}{\bibfnamefont{S.~A.}
  \bibnamefont{{Bonometto}}}, \bibinfo{journal}{\apjl}
  \textbf{\bibinfo{volume}{310}}, \bibinfo{pages}{L21} (\bibinfo{year}{1986}).

\bibitem[{\citenamefont{{Hwang} and {Noh}}(2005{\natexlab{a}})}]{HWNO05}
\bibinfo{author}{\bibfnamefont{J.-C.} \bibnamefont{{Hwang}}} \bibnamefont{and}
  \bibinfo{author}{\bibfnamefont{H.}~\bibnamefont{{Noh}}},
  \bibinfo{journal}{\prd} \textbf{\bibinfo{volume}{72}}, \bibinfo{eid}{044011}
  (\bibinfo{year}{2005}{\natexlab{a}}), \eprint{arXiv/0412128}.

\bibitem[{\citenamefont{{Hwang} and {Noh}}(2005{\natexlab{b}})}]{HWNO05b}
\bibinfo{author}{\bibfnamefont{J.-C.} \bibnamefont{{Hwang}}} \bibnamefont{and}
  \bibinfo{author}{\bibfnamefont{H.}~\bibnamefont{{Noh}}},
  \bibinfo{journal}{\prd} \textbf{\bibinfo{volume}{72}}, \bibinfo{eid}{044012}
  (\bibinfo{year}{2005}{\natexlab{b}}), \eprint{gr-qc/0412129}.

\bibitem[{\citenamefont{{Hwang} and {Noh}}(2007)}]{HWNO07b}
\bibinfo{author}{\bibfnamefont{J.-c.} \bibnamefont{{Hwang}}} \bibnamefont{and}
  \bibinfo{author}{\bibfnamefont{H.}~\bibnamefont{{Noh}}},
  \bibinfo{journal}{\jcap} \textbf{\bibinfo{volume}{12}}, \bibinfo{eid}{003}
  (\bibinfo{year}{2007}), \eprint{0704.2086}.

\bibitem[{\citenamefont{{Bertacca} et~al.}(2015)\citenamefont{{Bertacca},
  {Bartolo}, {Bruni}, {Koyama}, {Maartens}, {Matarrese}, {Sasaki}, and
  {Wands}}}]{BEBAET15}
\bibinfo{author}{\bibfnamefont{D.}~\bibnamefont{{Bertacca}}},
  \bibinfo{author}{\bibfnamefont{N.}~\bibnamefont{{Bartolo}}},
  \bibinfo{author}{\bibfnamefont{M.}~\bibnamefont{{Bruni}}},
  \bibinfo{author}{\bibfnamefont{K.}~\bibnamefont{{Koyama}}},
  \bibinfo{author}{\bibfnamefont{R.}~\bibnamefont{{Maartens}}},
  \bibinfo{author}{\bibfnamefont{S.}~\bibnamefont{{Matarrese}}},
  \bibinfo{author}{\bibfnamefont{M.}~\bibnamefont{{Sasaki}}}, \bibnamefont{and}
  \bibinfo{author}{\bibfnamefont{D.}~\bibnamefont{{Wands}}},
  \bibinfo{journal}{Classical and Quantum Gravity}
  \textbf{\bibinfo{volume}{32}}, \bibinfo{eid}{175019} (\bibinfo{year}{2015}),
  \eprint{1501.03163}.

\bibitem[{\citenamefont{{Yoo} and {Gong}}(2016)}]{YOGO16}
\bibinfo{author}{\bibfnamefont{J.}~\bibnamefont{{Yoo}}} \bibnamefont{and}
  \bibinfo{author}{\bibfnamefont{J.-O.} \bibnamefont{{Gong}}},
  \bibinfo{journal}{\jcap} \textbf{\bibinfo{volume}{7}}, \bibinfo{pages}{017}
  (\bibinfo{year}{2016}), \eprint{1602.06300}.

\bibitem[{\citenamefont{{Peebles}}(1980)}]{PEEBL80}
\bibinfo{author}{\bibfnamefont{P.~J.~E.} \bibnamefont{{Peebles}}},
  \emph{\bibinfo{title}{{The large-scale structure of the universe}}}
  (\bibinfo{publisher}{Princeton University Press, Princeton},
  \bibinfo{year}{1980}).

\bibitem[{\citenamefont{{Pajer} et~al.}(2013)\citenamefont{{Pajer}, {Schmidt},
  and {Zaldarriaga}}}]{PASCZA13}
\bibinfo{author}{\bibfnamefont{E.}~\bibnamefont{{Pajer}}},
  \bibinfo{author}{\bibfnamefont{F.}~\bibnamefont{{Schmidt}}},
  \bibnamefont{and}
  \bibinfo{author}{\bibfnamefont{M.}~\bibnamefont{{Zaldarriaga}}},
  \bibinfo{journal}{\prd} \textbf{\bibinfo{volume}{88}}, \bibinfo{eid}{083502}
  (\bibinfo{year}{2013}), \eprint{1305.0824}.

\bibitem[{\citenamefont{{Dai} et~al.}(2015{\natexlab{a}})\citenamefont{{Dai},
  {Pajer}, and {Schmidt}}}]{DAPASC15}
\bibinfo{author}{\bibfnamefont{L.}~\bibnamefont{{Dai}}},
  \bibinfo{author}{\bibfnamefont{E.}~\bibnamefont{{Pajer}}}, \bibnamefont{and}
  \bibinfo{author}{\bibfnamefont{F.}~\bibnamefont{{Schmidt}}},
  \bibinfo{journal}{ArXiv e-prints}  (\bibinfo{year}{2015}{\natexlab{a}}),
  \eprint{1504.00351}.

\bibitem[{\citenamefont{{Dai} et~al.}(2015{\natexlab{b}})\citenamefont{{Dai},
  {Pajer}, and {Schmidt}}}]{DAPASC15a}
\bibinfo{author}{\bibfnamefont{L.}~\bibnamefont{{Dai}}},
  \bibinfo{author}{\bibfnamefont{E.}~\bibnamefont{{Pajer}}}, \bibnamefont{and}
  \bibinfo{author}{\bibfnamefont{F.}~\bibnamefont{{Schmidt}}},
  \bibinfo{journal}{\jcap} \textbf{\bibinfo{volume}{2015}},
  \bibinfo{pages}{043} (\bibinfo{year}{2015}{\natexlab{b}}),
  \eprint{1502.02011}.

\bibitem[{\citenamefont{{Cabass} et~al.}(2017)\citenamefont{{Cabass}, {Pajer},
  and {Schmidt}}}]{CAPASC17}
\bibinfo{author}{\bibfnamefont{G.}~\bibnamefont{{Cabass}}},
  \bibinfo{author}{\bibfnamefont{E.}~\bibnamefont{{Pajer}}}, \bibnamefont{and}
  \bibinfo{author}{\bibfnamefont{F.}~\bibnamefont{{Schmidt}}},
  \bibinfo{journal}{\jcap} \textbf{\bibinfo{volume}{2017}}, \bibinfo{eid}{003}
  (\bibinfo{year}{2017}), \eprint{1612.00033}.

\bibitem[{\citenamefont{Ip and Schmidt}(2017)}]{IPSC17}
\bibinfo{author}{\bibfnamefont{H.~Y.} \bibnamefont{Ip}} \bibnamefont{and}
  \bibinfo{author}{\bibfnamefont{F.}~\bibnamefont{Schmidt}},
  \bibinfo{journal}{\jcap} \textbf{\bibinfo{volume}{02}}, \bibinfo{pages}{025}
  (\bibinfo{year}{2017}), \eprint{1610.01059}.

\bibitem[{\citenamefont{{Adamek}
  et~al.}(2016{\natexlab{a}})\citenamefont{{Adamek}, {Daverio}, {Durrer}, and
  {Kunz}}}]{ADDAET16}
\bibinfo{author}{\bibfnamefont{J.}~\bibnamefont{{Adamek}}},
  \bibinfo{author}{\bibfnamefont{D.}~\bibnamefont{{Daverio}}},
  \bibinfo{author}{\bibfnamefont{R.}~\bibnamefont{{Durrer}}}, \bibnamefont{and}
  \bibinfo{author}{\bibfnamefont{M.}~\bibnamefont{{Kunz}}},
  \bibinfo{journal}{Nature Physics} \textbf{\bibinfo{volume}{12}},
  \bibinfo{pages}{346} (\bibinfo{year}{2016}{\natexlab{a}}),
  \eprint{1509.01699}.

\bibitem[{\citenamefont{{Adamek}
  et~al.}(2016{\natexlab{b}})\citenamefont{{Adamek}, {Daverio}, {Durrer}, and
  {Kunz}}}]{ADDAET16b}
\bibinfo{author}{\bibfnamefont{J.}~\bibnamefont{{Adamek}}},
  \bibinfo{author}{\bibfnamefont{D.}~\bibnamefont{{Daverio}}},
  \bibinfo{author}{\bibfnamefont{R.}~\bibnamefont{{Durrer}}}, \bibnamefont{and}
  \bibinfo{author}{\bibfnamefont{M.}~\bibnamefont{{Kunz}}},
  \bibinfo{journal}{\jcap} \textbf{\bibinfo{volume}{2016}}, \bibinfo{eid}{053}
  (\bibinfo{year}{2016}{\natexlab{b}}), \eprint{1604.06065}.

\bibitem[{\citenamefont{{Borzyszkowski}
  et~al.}(2017)\citenamefont{{Borzyszkowski}, {Bertacca}, and
  {Porciani}}}]{BOBEPO17}
\bibinfo{author}{\bibfnamefont{M.}~\bibnamefont{{Borzyszkowski}}},
  \bibinfo{author}{\bibfnamefont{D.}~\bibnamefont{{Bertacca}}},
  \bibnamefont{and}
  \bibinfo{author}{\bibfnamefont{C.}~\bibnamefont{{Porciani}}},
  \bibinfo{journal}{\mnras} \textbf{\bibinfo{volume}{471}},
  \bibinfo{pages}{3899} (\bibinfo{year}{2017}), \eprint{1703.03407}.

\bibitem[{\citenamefont{{Lepori} et~al.}(2023)\citenamefont{{Lepori}, {Schulz},
  {Adamek}, and {Durrer}}}]{LESCET23}
\bibinfo{author}{\bibfnamefont{F.}~\bibnamefont{{Lepori}}},
  \bibinfo{author}{\bibfnamefont{S.}~\bibnamefont{{Schulz}}},
  \bibinfo{author}{\bibfnamefont{J.}~\bibnamefont{{Adamek}}}, \bibnamefont{and}
  \bibinfo{author}{\bibfnamefont{R.}~\bibnamefont{{Durrer}}},
  \bibinfo{journal}{\jcap} \textbf{\bibinfo{volume}{2023}}, \bibinfo{eid}{036}
  (\bibinfo{year}{2023}), \eprint{2209.10533}.

\bibitem[{\citenamefont{{Yoo} and {Zaldarriaga}}(2014)}]{YOZA14}
\bibinfo{author}{\bibfnamefont{J.}~\bibnamefont{{Yoo}}} \bibnamefont{and}
  \bibinfo{author}{\bibfnamefont{M.}~\bibnamefont{{Zaldarriaga}}},
  \bibinfo{journal}{\prd} \textbf{\bibinfo{volume}{90}},
  \bibinfo{pages}{023513} (\bibinfo{year}{2014}), \eprint{1406.4140}.

\bibitem[{\citenamefont{{Di Dio} et~al.}(2014)\citenamefont{{Di Dio}, {Durrer},
  {Marozzi}, and {Montanari}}}]{DIDUET14}
\bibinfo{author}{\bibfnamefont{E.}~\bibnamefont{{Di Dio}}},
  \bibinfo{author}{\bibfnamefont{R.}~\bibnamefont{{Durrer}}},
  \bibinfo{author}{\bibfnamefont{G.}~\bibnamefont{{Marozzi}}},
  \bibnamefont{and}
  \bibinfo{author}{\bibfnamefont{F.}~\bibnamefont{{Montanari}}},
  \bibinfo{journal}{\jcap} \textbf{\bibinfo{volume}{12}}, \bibinfo{eid}{017}
  (\bibinfo{year}{2014}), \eprint{1407.0376}.

\bibitem[{\citenamefont{{Bertacca}
  et~al.}(2014{\natexlab{a}})\citenamefont{{Bertacca}, {Maartens}, and
  {Clarkson}}}]{BEMACL14}
\bibinfo{author}{\bibfnamefont{D.}~\bibnamefont{{Bertacca}}},
  \bibinfo{author}{\bibfnamefont{R.}~\bibnamefont{{Maartens}}},
  \bibnamefont{and}
  \bibinfo{author}{\bibfnamefont{C.}~\bibnamefont{{Clarkson}}},
  \bibinfo{journal}{\jcap} \textbf{\bibinfo{volume}{9}}, \bibinfo{eid}{037}
  (\bibinfo{year}{2014}{\natexlab{a}}), \eprint{1405.4403}.

\bibitem[{\citenamefont{{Bertacca}
  et~al.}(2014{\natexlab{b}})\citenamefont{{Bertacca}, {Maartens}, and
  {Clarkson}}}]{BEMACL14b}
\bibinfo{author}{\bibfnamefont{D.}~\bibnamefont{{Bertacca}}},
  \bibinfo{author}{\bibfnamefont{R.}~\bibnamefont{{Maartens}}},
  \bibnamefont{and}
  \bibinfo{author}{\bibfnamefont{C.}~\bibnamefont{{Clarkson}}},
  \bibinfo{journal}{\jcap} \textbf{\bibinfo{volume}{11}}, \bibinfo{eid}{013}
  (\bibinfo{year}{2014}{\natexlab{b}}), \eprint{1406.0319}.

\bibitem[{\citenamefont{{Di Dio} et~al.}(2016)\citenamefont{{Di Dio}, {Durrer},
  {Marozzi}, and {Montanari}}}]{DIDUET16}
\bibinfo{author}{\bibfnamefont{E.}~\bibnamefont{{Di Dio}}},
  \bibinfo{author}{\bibfnamefont{R.}~\bibnamefont{{Durrer}}},
  \bibinfo{author}{\bibfnamefont{G.}~\bibnamefont{{Marozzi}}},
  \bibnamefont{and}
  \bibinfo{author}{\bibfnamefont{F.}~\bibnamefont{{Montanari}}},
  \bibinfo{journal}{\jcap} \textbf{\bibinfo{volume}{2016}},
  \bibinfo{pages}{016} (\bibinfo{year}{2016}), \eprint{1510.04202}.

\bibitem[{\citenamefont{{Umeh} et~al.}(2017)\citenamefont{{Umeh}, {Jolicoeur},
  {Maartens}, and {Clarkson}}}]{UMJOET17}
\bibinfo{author}{\bibfnamefont{O.}~\bibnamefont{{Umeh}}},
  \bibinfo{author}{\bibfnamefont{S.}~\bibnamefont{{Jolicoeur}}},
  \bibinfo{author}{\bibfnamefont{R.}~\bibnamefont{{Maartens}}},
  \bibnamefont{and}
  \bibinfo{author}{\bibfnamefont{C.}~\bibnamefont{{Clarkson}}},
  \bibinfo{journal}{\jcap} \textbf{\bibinfo{volume}{2017}}, \bibinfo{eid}{034}
  (\bibinfo{year}{2017}), \eprint{1610.03351}.

\bibitem[{\citenamefont{{Jolicoeur} et~al.}(2017)\citenamefont{{Jolicoeur},
  {Umeh}, {Maartens}, and {Clarkson}}}]{JOUMET17}
\bibinfo{author}{\bibfnamefont{S.}~\bibnamefont{{Jolicoeur}}},
  \bibinfo{author}{\bibfnamefont{O.}~\bibnamefont{{Umeh}}},
  \bibinfo{author}{\bibfnamefont{R.}~\bibnamefont{{Maartens}}},
  \bibnamefont{and}
  \bibinfo{author}{\bibfnamefont{C.}~\bibnamefont{{Clarkson}}},
  \bibinfo{journal}{\jcap} \textbf{\bibinfo{volume}{2017}}, \bibinfo{eid}{040}
  (\bibinfo{year}{2017}), \eprint{1703.09630}.

\bibitem[{\citenamefont{{Magi} and {Yoo}}(2022)}]{MAYO22}
\bibinfo{author}{\bibfnamefont{M.}~\bibnamefont{{Magi}}} \bibnamefont{and}
  \bibinfo{author}{\bibfnamefont{J.}~\bibnamefont{{Yoo}}},
  \bibinfo{journal}{\jcap} \textbf{\bibinfo{volume}{2022}}, \bibinfo{eid}{071}
  (\bibinfo{year}{2022}), \eprint{2204.01751}.

\bibitem[{\citenamefont{{Yoo}}(2009)}]{YOO09}
\bibinfo{author}{\bibfnamefont{J.}~\bibnamefont{{Yoo}}},
  \bibinfo{journal}{\prd} \textbf{\bibinfo{volume}{79}},
  \bibinfo{pages}{023517} (\bibinfo{year}{2009}), \eprint{arXiv:0808.3138}.

\bibitem[{\citenamefont{{Kaiser}}(1992)}]{KAISE92}
\bibinfo{author}{\bibfnamefont{N.}~\bibnamefont{{Kaiser}}},
  \bibinfo{journal}{\apj} \textbf{\bibinfo{volume}{388}}, \bibinfo{pages}{272}
  (\bibinfo{year}{1992}).

\bibitem[{\citenamefont{{Kaiser}}(1987)}]{KAISE87}
\bibinfo{author}{\bibfnamefont{N.}~\bibnamefont{{Kaiser}}},
  \bibinfo{journal}{\mnras} \textbf{\bibinfo{volume}{227}}, \bibinfo{pages}{1}
  (\bibinfo{year}{1987}).

\end{thebibliography}

\end{document}